\documentclass[sigconf, nonacm]{acmart}

\usepackage{epstopdf}
\usepackage{algorithmic}
\usepackage{graphicx}
\usepackage{textcomp}
\usepackage{fancyhdr}
\usepackage{color}
\usepackage{soul}
\usepackage{subfig}
\usepackage{epsfig}
\usepackage{alltt}
\usepackage{enumitem}
\usepackage[normalem]{ulem}
\usepackage{float}
\usepackage[]{algorithm2e}
\usepackage{lipsum}
\usepackage{calc}
\usepackage{multirow}
\usepackage{pifont}
\usepackage{adjustbox}
\usepackage{booktabs}
\usepackage{listings}
\usepackage{tikz}
\usepackage{xcolor, colortbl}
\fancypagestyle{specialfooter}{%
  \fancyhf{}
  
  \fancyfoot[C]{Response: Page \thepage\ of 2}
}

\usepackage{lipsum}
\usepackage{diagbox}
\usepackage{tabularx}
\usepackage[framemethod=tikz]{mdframed}
\usepackage{formatting/macros}
\usepackage{formatting/results}
\usepackage{algorithmic}

\usepackage{balance}

\usepackage{hyperref}
\usepackage{makecell}
\usepackage{color, soul}

\newcommand\vldbdoi{10.14778/3485450.3485462}
\newcommand\vldbpages{XXX-XXX}
\newcommand\vldbvolume{15}
\newcommand\vldbissue{1}
\newcommand\vldbyear{2022}
\newcommand\vldbauthors{\authors}
\newcommand\vldbtitle{\shorttitle} 
\newcommand\vldbavailabilityurl{empty}
\newcommand\vldbpagestyle{empty}

\begin{document}
\title{Accelerating Recommendation System Training \\ by Leveraging Popular Choices}

%%
%% The "author" command and its associated commands are used to define the authors and their affiliations.
\author{Muhammad Adnan}
\affiliation{%
  \institution{University of British Columbia}
%  \city{Vancouver}
%  \state{Canada}
}
\email{adnan@ece.ubc.ca}

\author{Yassaman Ebrahimzadeh Maboud}
%\orcid{0000-0002-1825-0097}
\affiliation{%
  \institution{University of British Columbia}
%  \city{Vancouver}
%  \state{Canada}
}
\email{yassaman@ece.ubc.ca}

\author{Divya Mahajan}
%\orcid{0000-0001-5109-3700}
\affiliation{%
  \institution{Microsoft}
%  \city{Redmond}
%  \country{USA}
}
\email{divya.mahajan@microsoft.com}

\author{Prashant J. Nair}
\affiliation{%
  \institution{University of British Columbia}
%  \city{Vancouver}
%  \state{Canada}
}
\email{prashantnair@ece.ubc.ca}

\begin{abstract}
%
%Scalable Machine-Learning training systems are the corner stone for improving user experience and model accuracy. 
%
%Training such large scale machine learning models often require extensive compute, memory, and bandwidth resources. 
%
Recommender models are commonly used to suggest relevant items to a user for e-commerce and online advertisement-based applications. 
These models use massive embedding tables to store numerical representation of items' and users' categorical variables (memory intensive) and employ neural networks (compute intensive) to generate final recommendations.
Training these large-scale recommendation models is evolving to require increasing data and compute resources.
%
%Although GPUs offer a means to accelerate the compute, they are limited by memory capacity, thus require the model data to be stored and transferred from the CPU. 
%
The highly parallel neural networks portion of these models can benefit from GPU acceleration however, large embedding tables often cannot fit in the limited-capacity GPU device memory.
Hence, this paper deep dives into the semantics of training data and obtains insights about the feature access, transfer, and usage patterns of these models. 
We observe that, due to the popularity of certain inputs, the accesses to the embeddings are highly skewed with a few embedding entries being accessed up to $10000\times$ more. 
This paper leverages this asymmetrical access pattern to offer a framework, called \framework, and proposes a hot-embedding aware data layout for training recommender models.
This layout utilizes the scarce GPU memory for storing the highly accessed embeddings, thus reduces the data transfers from CPU to GPU. 
At the same time, \framework engages the GPU to accelerate the executions of these hot embedding entries. 
Experiments on production-scale recommendation models with real datasets show that \framework reduces the overall training time by \improvementavgxdlcpu and \improvementavgxdlgpu in comparison to XDL CPU-only and XDL CPU-GPU execution while maintaining baseline accuracy.

\end{abstract}

\sloppy

\maketitle

\pagestyle{\vldbpagestyle}
\begingroup\small\noindent\raggedright\textbf{PVLDB Reference Format:}\\
\vldbauthors. \vldbtitle. PVLDB, \vldbvolume(\vldbissue): \vldbpages, \vldbyear.\\
\href{https://doi.org/\vldbdoi}{doi:\vldbdoi}
\endgroup
\begingroup
\renewcommand\thefootnote{}\footnote{\noindent
This work is licensed under the Creative Commons BY-NC-ND 4.0 International License. Visit \url{https://creativecommons.org/licenses/by-nc-nd/4.0/} to view a copy of this license. For any use beyond those covered by this license, obtain permission by emailing \href{mailto:info@vldb.org}{info@vldb.org}. Copyright is held by the owner/author(s). Publication rights licensed to the VLDB Endowment. \\
\raggedright Proceedings of the VLDB Endowment, Vol. \vldbvolume, No. \vldbissue\ %
ISSN 2150-8097. \\
\href{https://doi.org/\vldbdoi}{doi:\vldbdoi} \\
}\addtocounter{footnote}{-1}\endgroup

\ifdefempty{\vldbavailabilityurl}{}{
\vspace{.3cm}
\begingroup\small\noindent\raggedright\textbf{PVLDB Artifact Availability:}\\
The source code, data, and/or other artifacts have been made available at \url{https://github.com/Lab-Gattaca-UBC/Accelerating-RecSys-Training}.
\endgroup}
\setcounter{section}{0}
\setcounter{page}{1}
%------- PN Flow --------
% What are recommender systems and its challenges in learning machine learning.
% Organization of recommender systems : MLP Layers and Embedding Layers
% Characteristics of these layers -- 
% a. MLP layers -- highly parallel, low memory requirement and GPU memory friendly
% b Embedding Layer -- highly parallel, very high memory requirement and tends to be GPU memory unfriendly.
% In production, going into the future, the size of the embedding layers is only bound to increase. This presents a challenge.
% What options are present for tackling the issue of large embedding table sizes: GPU-GPU or CPU-GPU. GPU-GPU has communication overhead. However, CPU-GPU is advantageous as CPU can have very large memory capacities. --> MLP on GPU and Embedding Layer on CPU.

% Our Insight: Popular Data exist in the real world and these data samples access a set of hot entries in embedding tables. Present data to validate this insight.

%------------------------

\section{Introduction}
\label{sec:intro}

%
%Training large-scale machine-learning models has garnered widespread interest due to their compute and memory requirements. Proposals such as reduced-precision~\cite{binarizednns, limitednumericalprecision}, sparse~\cite{deepcompression}, and compressed~\cite{gist} data techniques are often used to reduce memory footprint and bandwidth requirements for these large-scale models. Furthermore, domain specific architectures are being rapidly developed to accelerate the compute~\cite{tpu, brainwave, dadiannao:micro:2014, tabla:hpca, eyeriss}. 

%At the same time, researchers are also investigating system-level distributed optimizations to increase the resource utilization on Graphical Processing Units (GPUs)~\cite{pipedream, gpipe}. 
%
%As these machine learning models become larger with billions of parameters to learn, training them has become more expensive and often requires multiple GPUs, i.e, distributed training..

%\todo{Divya}{Add citations here.} Machine learning models have garnered widespread interest due to their wide applicability in image processing~\cite{resnet}, object detection, healthcare, finance, and many more domains.
%

Recommendation models are an important class of machine learning algorithms that enable the industry (Netflix~\cite{netflixreco}, Facebook~\cite{dlrm}, Amazon~\cite{amazonreco}, etc.) to offer a targeted user experience through personalized recommendations. 
Deep learning based recommendation models~\cite{dlrm, tbsm} are at the core of a wide variety of internet services and consume significant infrastructure capacity and compute cycles in datacenters~\cite{fbtrainingreco}.
Training such at-scale models observes a conflation of challenges arising from high compute and data storage/transfer requirements. 
On the compute side, hardware accelerators notably GPUs and other heterogeneous architectures~\cite{tpu, brainwave, dadiannao:micro:2014, tabla:hpca, eyeriss} provide a robust mechanism to increase performance and energy efficiency.
To mitigate the large memory requirement, distributing training load through model parallel training~\cite{pipedream, gpipe, distributedtraining, largescalegputraining} or reducing the overall memory requirement through sparsity~\cite{deepcompression} and compression~\cite{hippogriff, compressreco, compressreco2, gist, touche,attache,thesaurus,dice} can be used. 
However, such techniques either require a pool of hardware accelerators that cumulatively provide enough memory to store these large models or tradeoff accuracy from the reduced precision for model footprint.

\vspace{-2ex}
\subsection{Motivation}
\label{sec:motivation}

\begin{figure*}[t]
\centering
	\includegraphics[width=0.95\textwidth]{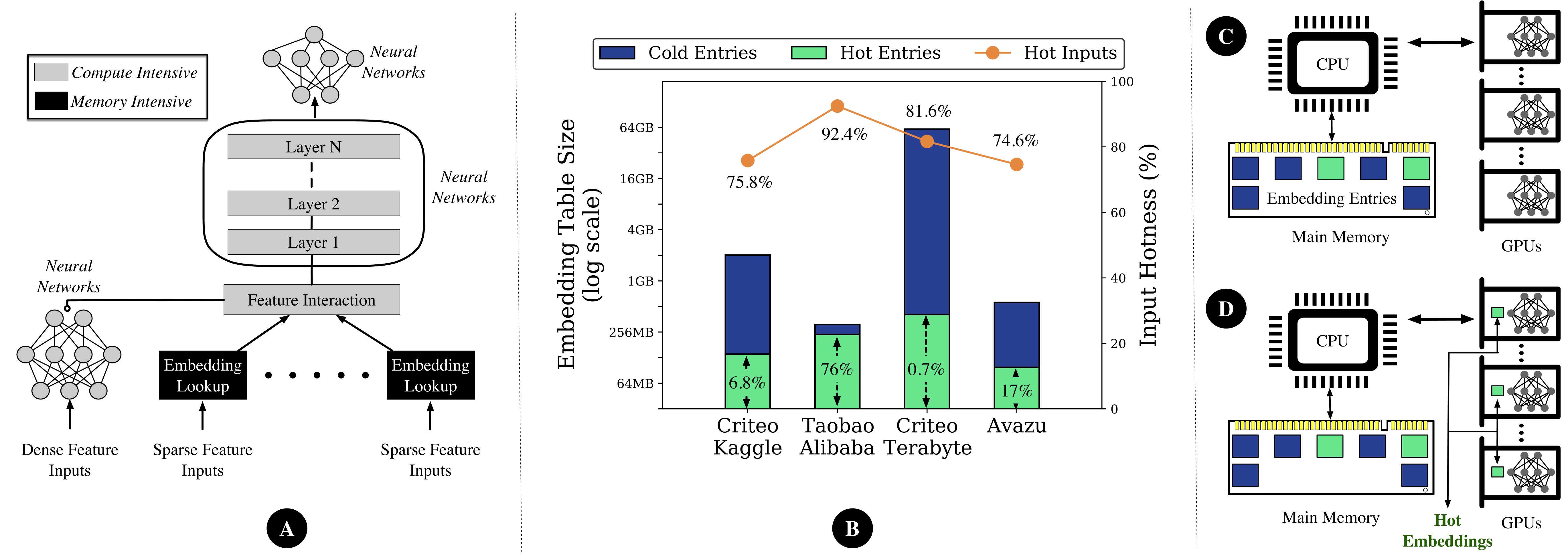}
	\caption{\encircle{A} Typical recommender model~\cite{dlrm, neuralcf, tbsm}. They comprise compute-intensive neural networks like DNNs and MLPs in tandem with the memory-intensive embedding tables. \encircle{B} shows embedding table sizes for four real world datasets and the proportion of the embedding table that is frequently accessed (hot). The graph also shows the \% of training inputs that only access the hot embeddings. \encircle{C} shows the baseline embedding data layout, i.e., storing entirely in the main memory. \encircle{D} shows the proposed layout where hot embeddings that cater to >70\% of the training inputs, are stored locally on GPUs.}
	\label{fig:recomodel}
\end{figure*}

Recommender models, as shown in Figure~\ref{fig:recomodel}A, use embedding tables that contribute heavily towards the memory capacity requirement and neural networks that exhibit compute intensity. 
While neural networks can benefit from GPUs, embedding tables (10s of GBs) often cannot fit within GPU memories~\cite{fbtrainingreco, yang2020mixedprecision,emergingmemacc}. 
Naively using model parallelism just to store the large embedding data across multiple GPUs is sub-optimal, as the number of GPU devices per compute node are not only fixed, but also scarce and expensive.
%
%Instead of using model parallelism across GPUs for memory capacity, they should ideally be used by large deep-learning models that also exhibit high computational intensity.
%
%Furthermore, at-scale recommendation models exhibit unique memory requirements per compute operation~\cite{facebookreco:hpca} compared to conventional deep models such as CNNs~\cite{resnet}, RNNs~\cite{tnlg}, and Transformers~\cite{bert}.

Figure~\ref{fig:recomodel}B shows the size of the embedding tables for \textit{four real-world datasets}~\cite{criteokaggle, alibaba, criteoterabyte, avazu} across two open-source recommender models, ``Deep Learning Recommendation Model for Personalization and Recommendation Systems'' (DLRM)~\cite{dlrm} and ``Time-based Sequence Model for Personalization and Recommendation Systems'' (TBSM)~\cite{tbsm}.
%
%We observe that, even on a high-end server class GPU device such as Nvidia Tesla-V100, recommendation models may not be able to fit the embedding tables (for some datasets) entirely within the GPU device memory.
%
As user-targeted applications evolve, the size of these embedding tables is expected to increase~\cite{tnlg, yang2020mixedprecision} at a rate faster than the anticipated increase in the memory capacity~\cite{mooreslaw, dark_silicon:babak}.
This is because larger embedding tables can track a greater and diverse degree of user preferences~\cite{fbtrainingreco}.
%
%Compression of data can also reduce the memory footprint but mainly reduces the communication overhead and requires decompression during runtime, whereas sparsity fundamentally changes the model and requires re-validation for accuracy. 
%
Therefore, in practice, it is common to train recommendation models either solely on CPUs or use the CPUs for handling the embedding data with GPUs executing data-parallel neural networks~\cite{aibox}. 
In the latter case, embeddings are stored in CPU memories as shown in Figure~\ref{fig:recomodel}C and require embedding data to be transferred between CPU and GPUs. 

Past work~\cite{energyproblem} has shown that data transfers not degrade performance but also consume significantly higher energy compared to accessing device memories.
To address this, we leverage the observation that certain embedding entries and inputs to recommendation models are significantly more popular than the others.
For instance, blockbuster movies tends to be significantly more popular than other movies.
Below, we discuss how popular inputs and embeddings can be delegated to a faster and compute-proximate device memory while maintaining the training fidelity.

\vspace{-2ex}
\subsection{Proposed Work and Contributions}

Prior work~\cite{wave,popsimilar} have shown that convergence of population preferences underlies the principle of popular inputs.
This popularity of training inputs implies that embedding data (accessed based on the input) also exhibits a highly skewed access behavior. 
Figure~\ref{fig:recomodel}B shows the portion of the embedding entries accessed by popular inputs in real-world datasets. 
For each benchmark, entries that have been accessed greater than $10^{-5}$\%, $10^{-6}$\%, $10^{-5}$\%,
$10^{-5}$\% of the total accesses respectively are showcased.
We call these highly accessed entries and their corresponding popular inputs as \emph{hot}.
This paper aims to offer an embedding data layout that accounts for access patterns of such models and their training inputs.
This data layout reduces the memory footprint of embedding data per GPU and mitigates frequent data transfers between CPU and GPUs.

\niparagraph{Optimized Data Layout:} 
The proposed optimized data layout classifies embedding entries into hot and cold regions as shown in Figure~\ref{fig:recomodel}D.
The categorization allows (1) replicating and storing \emph{only} the hot embedding data (only a few hundred MBs) on every GPU device memory and (2) perform all the hot embedding accesses and neural network tensor computations locally on the GPUs.
This eliminates any CPU-GPU communication for the popular inputs.  
For a large dataset like Criteo Terabyte, the size of \emph{hot} portions of embedding tables is about $\sim$400 MB (0.7\% as compared to 61GB for the entire tables) while catering to 81.6\% of the input data. 
These hot embeddings can easily fit within the memory of even a low-end GPUs.
\emph{For hot inputs, the entire graph shown in Figure~\ref{fig:recomodel} is trained using GPUs in a data-parallel fashion.
For the remainder of the inputs, their embedding accesses and computation are performed on the CPU and the neural network is executed in data parallel fashion on GPUs.}

\begin{figure*}[t]
	\centering
	\includegraphics[width=0.95\textwidth]{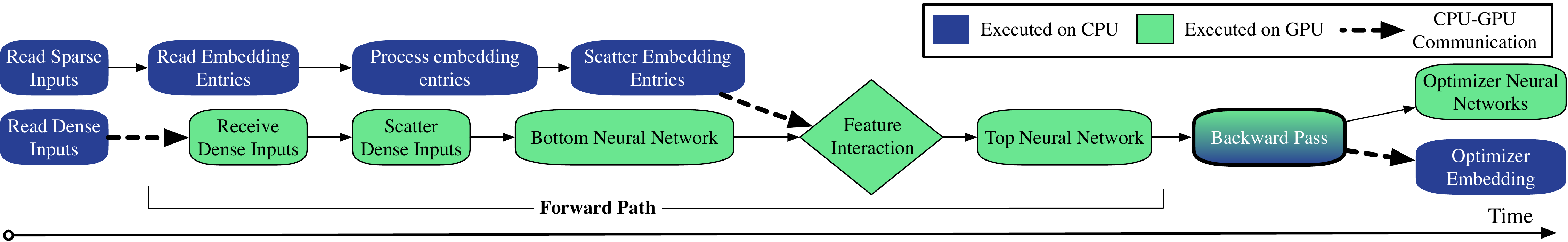}
	\caption{Execution graph of deep learning based recommender model. In this graph we show the forward graph in detail, the backward pass is a mirror of forward and executes on CPU and GPU according to its forward counterpart. The current mode of training for DLRM and TBSM requires embedding storage, reading, and processing, on CPU.}
	\label{fig:recograph}
\end{figure*}

\niparagraph{Challenges:} 
Storing hot embedding data locally \underline{in} every GPU poses four challenges:
First, as each training step executes a mini-batch of inputs. 
If even a single input within the mini-batch accesses a cold embedding entry, that data has to be obtained from the CPU. Thus incurs a CPU-GPU communication overhead and becomes the latency bottleneck for that mini-batch. 
Second, training contiguously \textit{only} on either hot or cold inputs can have an impact on accuracy. 
This is because, popular inputs only update the hot embeddings.
Third, as we split hot and cold embedding data between CPU and GPUs, all the devices need to be kept synchronized.
Fourth, the hotness of an embedding entry depends on the dataset and recommender model. 
Hence, hotness needs to be re-calibrated for every (model, dataset, and system configuration) tuple.

\niparagraph{Contributions:}
\emph{This paper proposes the \textit{Frequently Accessed Embeddings} (\framework) framework that efficiently places embedding data across CPUs and GPUs. while maintaining baseline accuracy}. This paper makes the following contributions:

%The \framework framework improves the performance of recommender model training while maintaining accuracy.} 
%
%
%\framework does not change the model or its training and exploits a runtime data access pattern to improve higher performance and makes the following contributions:

\begin{enumerate}
    \vspace{-1ex}
    \item We find that embedding table accesses in real world recommender models is heavily skewed, thus allocating equal compute resources to all the entries is sub-optimal. 
    \item We intelligently place hot embeddings on every GPU device involved in training while retaining cold entries on CPUs. 
    Placing only hot embeddings on GPUs reduces its memory requirement and improves performance. 
    This is because \framework eliminates CPU-GPU communication for inputs that access hot embeddings and enables accelerating the compute that involves those entries.
    %. % \framework determines which embedding entries are hot and cold and dynamically schedules inputs to ensure training accuracy.
    \item To optimize training, \framework performs sampling of the input dataset to determine the access pattern of embedding tables. 
    Thereafter, \framework classifies the input data into hot and cold categories.
    \framework ensures that a mini-batch either accesses only hot or only cold embeddings to avoid communication overheads. 
    At runtime, \framework intertwines executions of hot and cold input mini-batches to ensure the baseline accuracy.
    \item \framework employs statistical techniques to avoid traversing through the entire input dataset and embedding tables to determine the hot embedding access threshold and the size of the hot embedding table while incurring negligible overhead.
\end{enumerate}

We prototype \framework on open-source deep learning-based recommender system training models DLRM~\cite{dlrm} and TBSM~\cite{tbsm}. These models are adopted by both academia~\cite{tensorcasting} and industry~\cite{facebook:ml, facebookreco:hpca, nvidiareco}.
We compare our \framework optimized training with two implementations. First, the open-source implementations of DLRM and TBSM. Second, a highly optimized implementation of these models using the \emph{XDL} framework~\cite{xdl}. 
We evaluate \framework for a wide variety of real-world and synthetic deep learning based recommender models.
% 
%We compare \framework optimized training with two baselines, first, where the entire training happens on server class CPU and second, where the embedding data is naively stored and processed on the CPU and the neural network on GPU.
%
For real-world model architectures, our experiments show that \framework achieves an average speedup of \improvementavgxdlcpu and \improvementavgxdlgpu in comparison to XDL enhanced CPU and CPU-GPU baseline, respectively.
Furthermore, \framework achieves  \improvementavgcpu and \improvementavgfourgpu against DLRM and TBSM implementations on CPU and CPU-GPU, respectively.
Both baselines execute in a mode that uses a CPU with 4 GPUs.
\framework reduces the amount of data transferred from CPU to GPU by \avgdatatransferreducionxdl in comparison to XDL-based baseline.
%\
For synthetic model architectures, \framework achieves \synimprovementavgfourgpu speedup over XDL-based baseline.
\section{Background}
\label{sec:background}

In this section we provide the background on the model, inputs, and training process of recommendation systems.

\niparagraph{Recommendation models and their training inputs:}
Figure~\ref{fig:recograph} shows the flow of a recommendation model which comprises embedding lookup and neural network layers. 
The recommendation model has two types of inputs, namely sparse and dense. 
Sparse inputs typically denote specific preferences of the user (like the movie genre, choice of music, etc.) and are used by the embedding layers.
Dense inputs are continuous inputs (such as time of day, location of users, etc.) that feed directly into the neural network layers. 
The embedding phase uses large tables containing data that reduces the sparse input feature space into a vector.
%
%Operations involving embedding tables are highly memory intensive as these tables are of the order of 10's of GBs. 
%
These inputs are used by the Deep Neural Network (DNN) and Multi-Layer Perceptron (MLP) components to classify and determine the final recommendation.

\niparagraph{State-of-the-art mode of execution for training.}
Machine learning techniques generally employ data-parallel training to reduce the overall execution time~\cite{dataparallel}. 
This mode of training requires model replication across all the GPU devices, where each device executes on different inputs in a mini-batch. 
Thereafter, a post-execution synchronization is performed to update the weights/parameters using the aggregated gradient values. 
%
%Other common forms of training include model parallel~\cite{modelparallel} and hybrid data and model parallel~\cite{hybridmodeldata1, hybridmodeldata2, pipedream} modes based on the requirements of the model and the system limitations.
%
%Often model parallel is employed to support training of very large models which cannot fit on a single device. 
%
For recommendation models, this training mode tends to be infeasible as embedding tables cannot fit even on high-end GPUs such as Nvidia-V100.
%
%However for recommendation models, as mentioned above, embedding tables constitute the major portion of the memory requirements. 
%
%Even beyond recommendation models, large embeddings are being used to represent categorical features~\cite{tnlg, bert} across various models.

To overcome this issue, as shown in the Figure~\ref{fig:recograph}, past work either executes the whole graph on the CPU or uses the CPU to handle the memory-intensive embedding layer with the GPUs executing the compute-intensive DNN layers.
The first case is inefficient as CPUs are not optimized for neural network training as they cannot optimally process large tensor operations.
%
%This further implies that recommender models cannot leverage the accelerated compute in GPUs. 
%
On the other hand, the hybrid CPU-GPU mode incurs CPU-GPU communication overheads for intermediate results and gradients. 
This is shown in the forward pass by the bold dotted lines in the Figure~\ref{fig:recograph}.
%
%In this case, the embedding processing happens on the CPU and is then sent to the GPU and consumed by the neural network layers.
%
The backward pass also executes in a CPU-GPU mode, with CPU executing the backward computation for embeddings and GPU executing the backward propagation of neural layers. 
Thereafter, the gradients are generated on CPU for embeddings and on GPU for neural layers. 
Our experiments show that CPU-GPU communication can take up to \cpugpucommunicationoverhead of the total training time. 
Additionally, any computation involving embedding data, such as the massively-parallel Stochastic Gradient Descent optimization, also then executes on the CPU. 
%
%In conclusion, storing of large embeddings on CPU leads to data being transferred between CPU and GPU. Furthermore, CPU also incurs additional compute for processing these embeddings. 

%As the embedding layer is stored on the CPU, after the backward pass, the training optimizer , which is a massively parallel operation, is also performed on the CPU. 
%
%Alternatively, in the second technique, embedding tables are split across multiple GPUs (model parallelism)~\cite{} whereas the neural network portion executes in a data parallel mode as shown in Figure~XXXXX.
%
%This form of training incurs a high latency butterfly shuffle as the output from the embedding layer is scattered to every GPU as the inputs to the neural network layer.
%
%There are optimizations that are deployed~\cite{blink, fbquantcollective} to reduce the overhead of the collective calls such as butterfly shuffle and scatter, but they are still limited by the underlying network support. 
%
%Even on high-end servers, where GPUs are connected using dedicated NVLink~\cite{nvlink} interconnects, communication overheads often present a performance-power concern.

\niparagraph{Leveraging training input and embedding access patterns:}
Data accesses can exhibit locality that can be exploited either at software~\cite{mrshare, hippogriff}, system~\cite{quiver}, or hardware~\cite{dana} level.
For recommender models trained on real-world data, some sparse inputs are significantly more popular than others.
Therefore, in \textit{such real-world applications}, accesses into embedding tables are heavily skewed.
For instance, for the Criteo Kaggle dataset~\cite{criteokaggle} on DLRM, the top $6.8$\% of the embedding table entries observe at least $76$\% of the total accesses.
\emph{It is important to note that the cold portion of the embedding data is critical from a learning perspective as it contributes to the accuracy of the model. 
Training only on popular inputs would make the targeted user experience futile as it would lead to certain items being always recommended.}
Nevertheless, from a memory perspective, as shown in Figure~\ref{fig:recomodel}B, hot entries are more important as they form \hotentriespercentageaccesslow to \hotentriespercentageaccesshigh of the total training input accesses.

%
%To establish: an embedding entry is hot if it captures at least the specified percentage of the total accesses and a hot input is one that only accesses these hot entries.
%
%However, the above mentioned CPU-only and CPU-GPU hybrid approach naively assumes that all the embedding data is equally important in terms of resource requirement. 
This paper leverages the popularity semantics of training input to mitigate the bottlenecks of the above mentioned CPU-GPU execution by optimizing the embedding data layout in the memory hierarchy. 
\emph{Intuitively, highly accessed embeddings are kept in close proximity to the compute, i.e. GPU, whereas the cold embedding entries are stored in relatively larger but slower CPU memories.}
%
%Keeping the data parallel training process in mind, this work replicates the hot embedding entries on every GPU.
%
This allows us to execute the entire training graph, shown in Figure~\ref{fig:recograph}, on the GPU in a data-parallel fashion for the popular inputs.
%
%As we show in Figure~\ref{fig:embhotminibatch}, these hot inputs form a major portion of the total training dataset.
%
This data layout overcomes the limitations of the baseline by - (1) \emph{accelerating embedding compute} through GPUs whilst being \emph{within the memory capacity} of the device and (2) \emph{eliminating the communication overheads} (gradients and activations) between CPU and GPU.
%
%Moreover, we train in pure data parallel mode for the hot embedding executions, instead of model and data parallel hybrid mode, thus circumventing part of the GPU to GPU (butterfly shuffle) overheads.
%

\section{Challenges and Insights}
\label{sec:challenges}

%The goal of this work is to offer a framework that can exploit the hot-embeddings and popular-input insight to reduce the overall training time for recommender systems.
%
To perform efficient end-to-end training with the optimized embedding layout while maintaining baseline accuracy, we require a comprehensive framework that has both static and runtime components. 
Next we analyze the challenges of such a training execution. 
%
%Below we discuss the insights that guide the design of our framework.

\begin{figure}[h!]
	\centering
	\includegraphics[width=1\columnwidth]{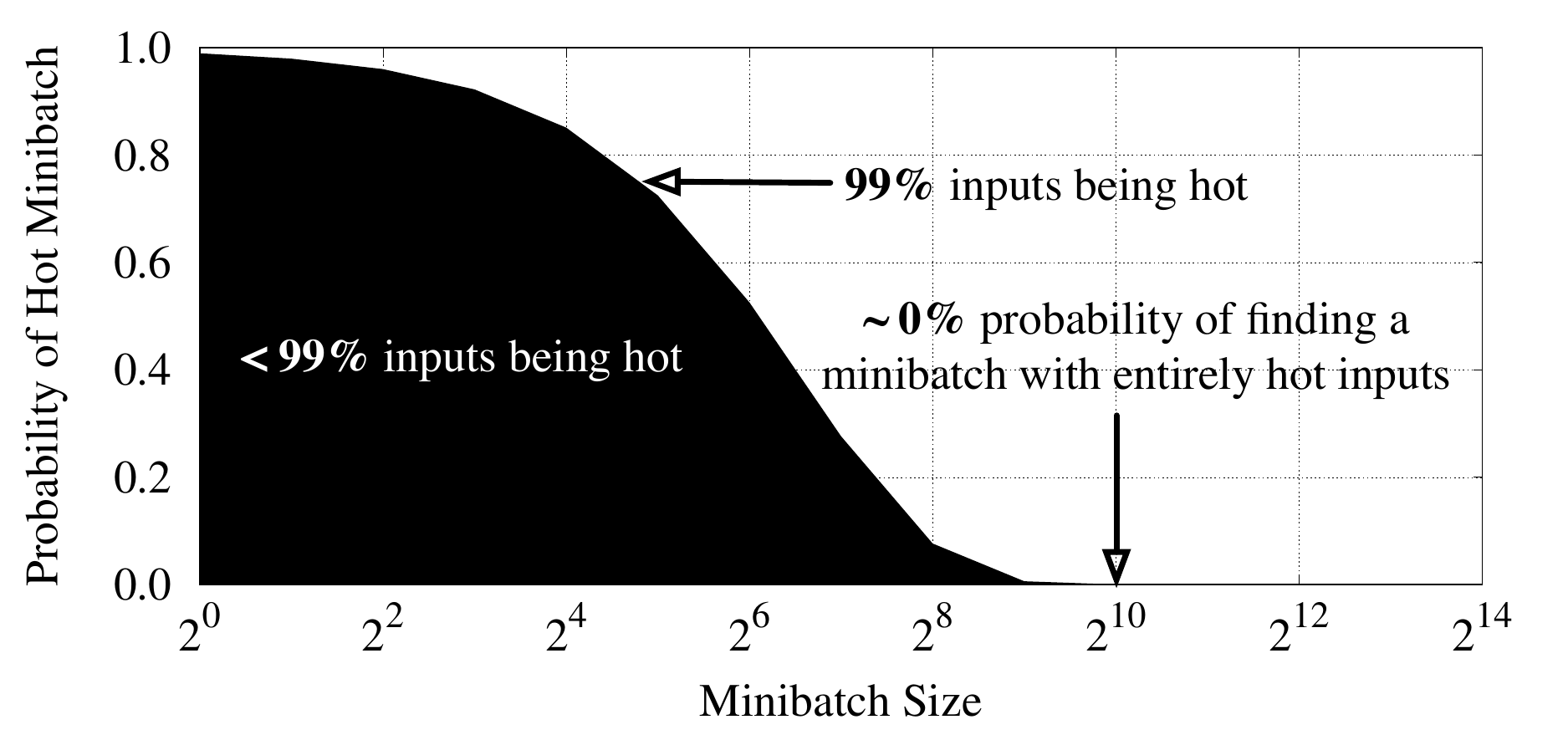}
	\caption{Probability of creating a mini-batch with all popular inputs when the number of hot-inputs is 99\% or lower. This reduces drastically as the mini-batch size increases.} %as it is likely that at least one input in a large mini-batch is not hot and therefore classifies the entire mini-batch as not-hot.}
	\label{fig:embhotminibatch}
\end{figure}

\niparagraph{(1) Does moving the hot embedding data to the GPU suffice?}
%
%Machine learning training uses a mini-batch of inputs distributed across machines in a data parallel fashion.
%
As shown in Figure~\ref{fig:embhotminibatch}, even if 99\% of the inputs are popular, i.e., access hot embeddings, the probability that the entire mini-batch accesses \textit{only} hot embeddings decreases dramatically as the minibatch size increases. 
This is because, it is likely that at least one input within a large mini-batch would require accessing cold embedding entries.
To obtain benefits from embedding data layout, we require the entire mini-batch to only access hot embedding entries. 
Even a single input accessing cold embedding entries would stall GPU execution as it tries to obtain its embedding entries from the CPU memory.
To overcome this challenge, our framework comprises a static component that performs input-dataset pre-processing and organizes mini-batches such that they completely contain only hot or cold inputs. 
This pre-processing needs to be performed \textit{only} once per dataset and is stored in a pre-processed format for subsequent executions.
For hot mini-batches, the framework performs GPU-only data-parallel execution and for cold mini-batches the framework falls back onto the CPU-GPU hybrid mode.

\niparagraph{(2) What constitutes a hot embedding entry?}
The classification of an embedding entry as hot or cold is based on the access threshold. 
Any entry that is accessed more than a threshold is classified as hot. 
We expose this threshold as a knob to \framework to adjust the amount of hot embeddings that can be managed by GPUs, based on both the model and system specifications.
To minimize performance overhead, we devise statistical techniques that use input dataset sampling to determine the access threshold.
This enables \framework to determine the optimal threshold without scanning the entire training data.
\framework selects a threshold that classifies enough embedding entries as hot so that they fits in allocated GPU device memory. 
%to obtain the access pattern of the embedding table.
%
%\framework analyzes only the accesses of the sampled input dataset to estimate the size of hot embedding table.
%
%The overarching goal of using a caching mechanism to store the hot regions is to reduce the memory footprint on the GPU device, as storing the entire model on GPU can accelerate the compute and reduce communication overheads. 
%
%Thus, we devise an algorithm to determine the threshold that in turn decides whether an entry belongs to a hot embedding table or not based on the GPU memory, recommender model, and the input dataset.  

\begin{figure*}
	\centering
	\includegraphics[width=0.95\textwidth]{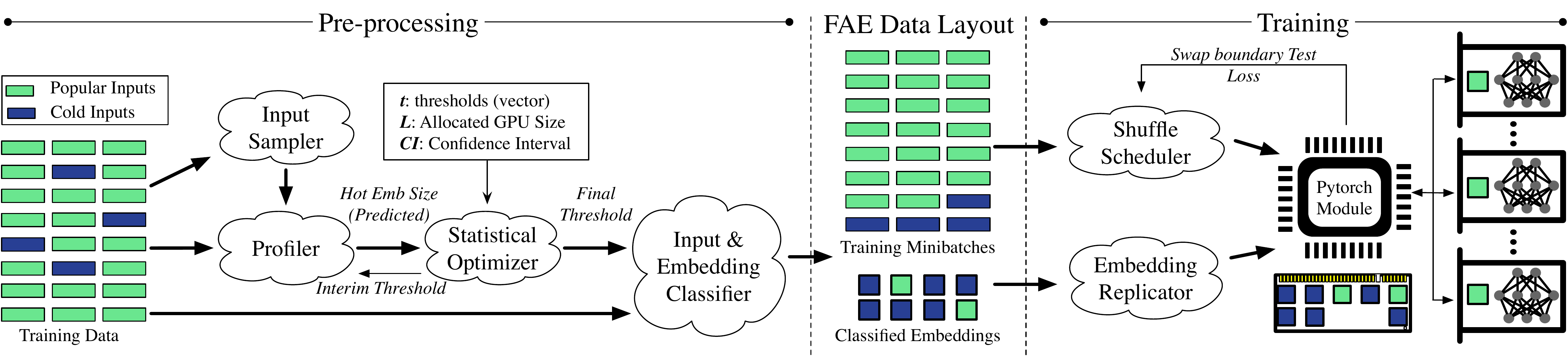}
	\caption{The \framework framework. The pre-processing phase calculates the threshold for classifying hot embeddings. This phase uses random-sampling of input datasets and embedding tables to determine the best threshold for hot embeddings. This threshold is also used to classify inputs into hot and cold mini-batches. At runtime, GPUs execute the hot input mini-batch while cold inputs execute in a CPU-GPU hybrid mode. The \scheduler uses feedback from the pytorch modules to determine the rate of hot and cold mini-batches swap.}
	\label{fig:fae}
\end{figure*}

\niparagraph{(3) How to schedule hot and cold mini-batches?}
\framework processed data contains mini-batches that are either entirely hot or cold.
%
%As mentioned above transitioning their executions incurs the embedding update overhead.
%
Scheduling all the hot mini-batches followed by cold mini-batches incurs the least embedding update overhead as the embeddings only have to synchronized between GPU and CPU once after the swap. %, as there are only a few hot entries.
However, such a technique can can have an non-negligible impact on the accuracy. 
This is because the hot mini-batches only update the hot embedding entries whereas the cold mini-batches cover more embedding entries (both hot and cold), albeit sparsely. 
%
%Even though the total number of times an embedding entry is updated does not vary the sequence in which it is updated can have an impact on accuracy. 
%
To tackle this issue, our framework, offers a runtime solution that dynamically tunes the rate of issuing hot and cold mini-batches to ensure that the accuracy metrics are met.
%
%In the next section, we discuss in detail the design decisions we made to devise such a framework.

%
%Additionally, these hot regions are also available with the entire embedding table either on the CPU or the distributed model parallel GPU form to ensure the baseline execution for the minibatches that do not contain only hot inputs. 
%
%This implies for the hot input mini-batch, where both embedding and MLP execute in a data parallel mode, all the copies of the tables have to be updates. 
%
%Moreover, now the optimizer for both the embedding and MLP also executes on the GPU-, thus these gradients do have to be send between devices. 
%

\niparagraph{(4) How to maintain consistency between the embedding tables that are scattered across devices?}
\framework replicates hot embedding tables across all the GPU devices and CPU contains all the embeddings (including hot embeddings). 
%
%This is because CPU executes cold inputs (cold inputs might access some hot embedding entries).
%
Thus, we need to perform two forms of synchronization during the training - one across all the GPUs after each mini-batch of data parallel execution and once between the cold and hot swap between CPU and GPU.
In the former case, hot embeddings are synchronized using the AllReduce collectives over the fast NVlink GPU to GPU interconnect~\cite{nvlink}.
%
%When moving execution from hot to cold mini-batches, or vice-versa, the hot embedding entries are updated on the CPU and GPU. 
%
%The non-hot inputs update the embeddings in the baseline mode.
%
%FAE increases the size of the gradient (additional hot embedding data) that needs to be synchronized, thus amortizes the collectives setup cost across neural model and embeddings, but incurs the raw communication cost over the high-bandwidth NVlink.
%
In the latter case, the synchronization across GPU and CPU between hot and cold mini-batches is performed through PCIe transfer between the GPU-CPU devices. 
This communication overhead incurred by \framework is accounted for in the final execution latencies. 
To reduce this overhead, \framework minimizes the transitions between hot and cold mini-batches, without compromising baseline accuracy.

%Using these insights, we design a framework that not only reduces the training time but also minimizes the communication load, energy consumption, and maintains baseline accuracy.
\section{The \framework Framework}
\label{sec:design}

This paper proposes the \textit{Frequently Accessed Embeddings} (\framework) framework to accelerate recommender system training.
\framework efficiently utilizes the GPU memory and computation throughput to reduce the communication cost of obtaining embedding data. 
Figure~\ref{fig:fae} illustrates the flow of the framework; \framework consists of (1) the input and embedding pre-processing stage that determines the hotness of embeddings by sampling the input training data and (2) the training stage that replicates hot embeddings on all the GPUs and schedules hot/cold minibatches to ensure baseline accuracy. 
%
%four components to efficiently capture, manage, and place hot embedding data, replicating it across GPUs, while dynamically calibrating the accuracy and communication overheads.
%
The pre-processing phase converges on an access threshold to classify an embedding entry as hot. 
This threshold is based on the allocated GPU memory size, confidence interval, and the CPU-GPU bandwidth.
Thereafter, based on the final threshold, the \emph{\embclassifier} and \emph{\inputclassifier} categorize both embedding entries and sparse inputs into hot and cold portions. 
%
%Hot inputs only access hot embeddings whereas cold inputs can access both hot or cold embedding entries.
%
The pre-processing phase executes statically \textit{once} per training dataset, and stores the pre-processed data in the \framework format for subsequent training runs.
At runtime, the \emph{\embreplicator}, extracts \emph{hot} embedding entries and creates embedding bags that are replicated across GPUs.
The \emph{\scheduler} dynamically determines the execution order of hot and cold sparse input mini-batches across the CPU and GPUs at runtime. 
Based on accuracy goals, the \scheduler interleaves hot and cold mini-batch queues to capture the updates to all embedding table entries.
To help understand the next few sub-sections, Table~\ref{table:notations} provides description of the notations for the design variables in \framework.
%This ensures that baseline accuracy requirements are met.

%Ideally, one would like the \calibrator to declare a large number of inputs to be classified as hot inputs. However, this would directly increase the size of the hot embedding bag and cause higher GPU memory overheads. 

\begin{table}[h!]
\centering
\caption{List of Notations}
\resizebox{1\columnwidth}{!}{
\begin{tabular}{c|l}
 \hline
 \textbf{Notation} & \textbf{Description} \\
 \hline \hline
 \textbf{\emph{D}} & Training input dataset\\
 \textbf{\emph{t}} & Minimum number of access to classify an entry as hot\\
 %\hline
 \textbf{\emph{T}} & Total number of accesses into an embedding table\\
  %\hline
 \textbf{\emph{L}} & User-specified allocation of GPU memory for hot embeddings\\
 %\hline
 \textbf{\emph{h}} & Maximum number of hot embeddings that fit in \textbf{\emph{L}}\\
 \textbf{\emph{E$_{z}$}} & Size of embedding table number \textbf{\emph{z}}\\
 \hline
 \textbf{\emph{x}} & Sampling rate for inputs (\%) \\
 %\hline
  $\widehat{\textbf{\emph{D}}}$ & \emph{Sampled} training input dataset entries\\
 \textbf{\emph{n}} & Number of Sample Chunks from the embedding logger \\
  %\hline
 \textbf{\emph{m}} & Number of entries in each embedding logger chunk (\emph{n})\\
 %\hline
 \textbf{\emph{N}} & Total \textbf{\emph{m}}-sized entries in the embedding logger\\
 %\hline
 \textbf{\emph{k}} & For any \textbf{\emph{t}} $\longrightarrow$ Total accesses into any embedding entry\\
 %\hline
 \textbf{\emph{H}}$_{zt}$ & For any \textbf{\emph{t}} $\longrightarrow$ Sample adjusted \emph{t} per (\emph{z}); minimum accesses to classify hot entries\\
 %\hline
\textbf{\emph{C}}$_{i}$ & For any \textbf{\emph{t}} $\longrightarrow$ Number of entries in the \emph{m} chunk with accesses more than \emph{H}$_{zt}$ \\
%\hline
$\bar{\textbf{\emph{y}}}$ & For any \textbf{\emph{t}} $\longrightarrow$ Mean of \textbf{\emph{C}}\\
%\hline
\textbf{\emph{s}} & For any \textbf{\emph{t}} $\longrightarrow$ Standard deviation of \textbf{\emph{C}}\\
%\hline
\textbf{\emph{CI}$_{\beta}$} & Confidence Interval of \textbf{$\beta$}\%\\
 \hline
\end{tabular}
}
\label{table:notations}
\end{table}

\subsection{Calibrating the Access Threshold}
\label{sec:threshold}

\begin{figure}[h!]
	\centering
	\includegraphics[width=1\columnwidth]{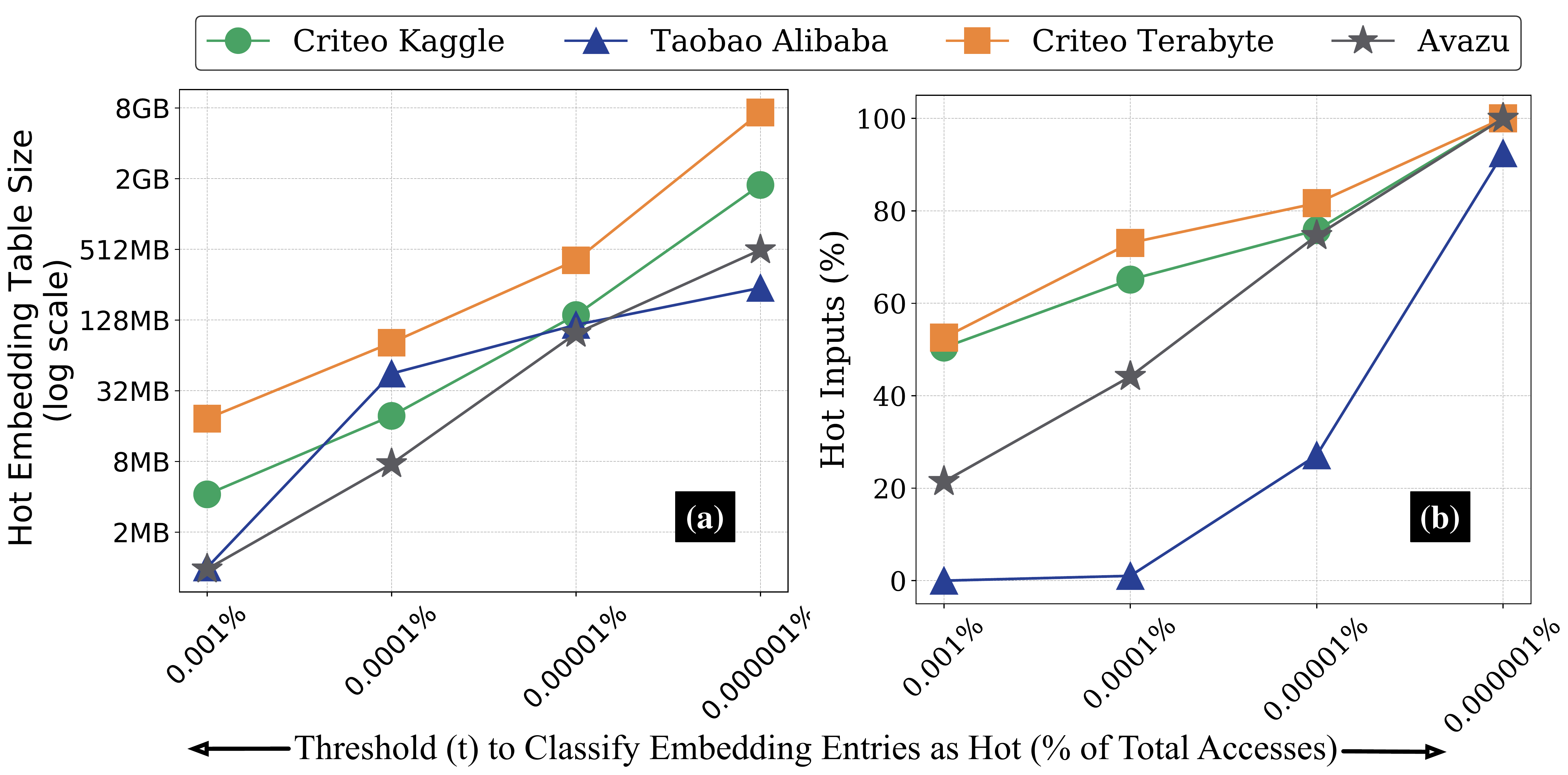}
	\caption{(a) Size of hot embedding entries and (b) Percentage of hot inputs with varying access threshold values. As we vary the threshold, the size of the embedding entries increases more rapidly compared to the percent of hot inputs}.
	\label{fig:hotinputembsize}
\end{figure}

The first goal of the pre-processing phase is to pick an access threshold (\textbf{\emph{t}}) for the embedding entries.
%
%The threshold determine tradeoff the size of the hot embeddings and performance. 
%
%any embedding entry accessed} $\geq$ \textbf{\emph{t}}$\times$\textbf{\emph{T}} 
%
%\revision{is considered hot.}
We denote \textbf{\emph{T}} as the total number of accesses into an embedding table. The accesses per entry for hot embeddings is $\geq$ \textbf{\emph{t}}$\times$\textbf{\emph{T}}.
Any input that accesses only hot embeddings is also categorized as hot.
%
%A system can have several 100s of GBs of Dynamic Random Access Memory (DRAM) on CPU. 
%
%For instance, even low-end CPUs like Intel-Core support up to 128GB DRAM capacity. 
%
%As high-end GPU devices support only 10s of GBs of DRAM memory, for example high-end Nvdia Tesla-V100 supports up to 32GB of DRAM, therefore, it is vital to ensure that the GPU memory is efficiently utilized. 
%
Picking a large \textbf{\emph{t}} would imply that only a few embedding entries would have enough accesses to be classified as hot.
It would lead to only a small percentage of sparse-inputs that would execute completely in a GPU execution mode and thus reduce the overall performance benefits.
Conversely, picking a small threshold will categorize embedding entries with very few accesses as hot which, would increase the embedding table size, often beyond the GPU device memory capacity.
Figure~\ref{fig:hotinputembsize} shows that we observe diminishing returns by reducing the threshold, as the number of hot embedding entries increases more steeply as compared to hot inputs.
%
%Moreover, as we later show in Figure~\ref{fig:inputprocessor}, shows the latency to process and categorize the input data with varying threshold.
%
%As the threshold is decreased, the latency to pre-process sparse-inputs also increases as a larger set of inputs would now scan a greater amount of embedding table entries.
%
Thus, we need to efficiently tune \textbf{\emph{t}} based on the system configuration parameters.

\begin{figure*}
	\centering
	\includegraphics[width=0.9\textwidth]{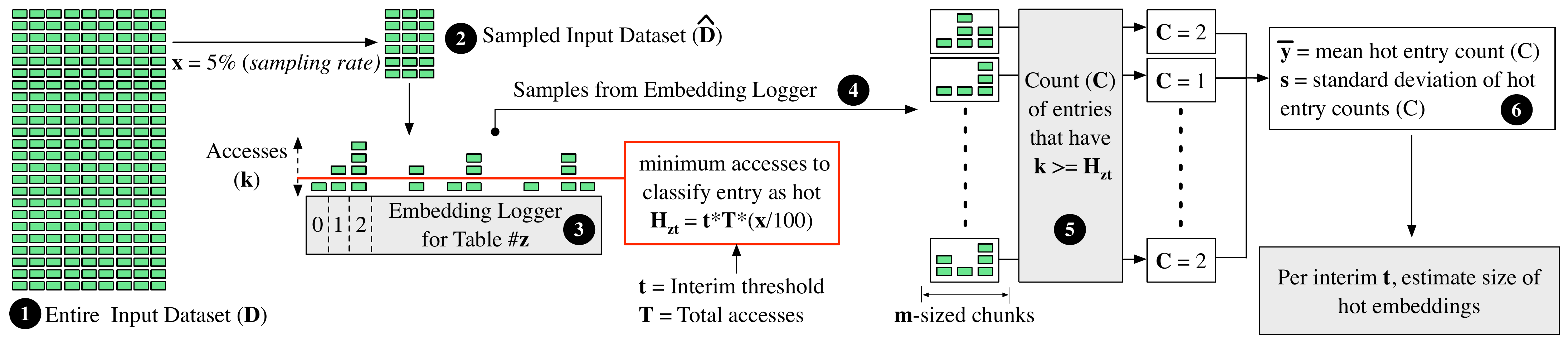}
	\caption{The flow of events in Input Sampler and Profiler. The original input \encircle{1} is sampled \encircle{2} at 5\%. This sample is used by the \emph{profiler} to create an access profile across embedding entries in the logger \encircle{3}. For each threshold,  A few chunks from the embedding logger are randomly sampled \encircle{4} to estimate the count of hot entries \encircle{5}. The mean and standard deviation of this count determines the size of hot embedding tables per threshold \encircle{6}.}
	\label{fig:preprocessingflow}
\end{figure*}

One of the system configuration parameters is the GPU memory allocated for hot embeddings -- denoted by \textbf{\emph{L}}.
Notation \textbf{\emph{h}} constitutes the maximum number of hot entries that fit within \textbf{\emph{L}}.
A naive mechanism to determine \textbf{\emph{t}} will profile the \emph{entire} training dataset and analyze the accesses of \emph{all} the embedding entries.
This requires sorting all embedding entries based on their access frequencies and classifying the top \textbf{\emph{h}} entries as hot. 
This implementation will incur a high pre-processing overhead as it could imply processing several terabytes of data -- even though profiling is performed only once per dataset.
%
%This is because the profiler must access the entire dataset and all embeddings, .} 
%
%
%As discussed earlier, hot inputs can be executed in an accelerated data-parallel manner on GPUs.
%
Instead, we propose a novel \emph{input sampler} and \emph{\statoptimizer} that ensures a low static compilation overhead for finding optimal \textbf{\emph{t}} such that \textbf{\emph{L}} is used effectively. Figure~\ref{fig:preprocessingflow} describes the flow of events to determine the optimal value of \textbf{\emph{t}}.

\subsubsection{Mitigating Read Overheads with Sparse Input Sampler}
\begin{figure}[!h]
	\centering
	\includegraphics[width=1\columnwidth]{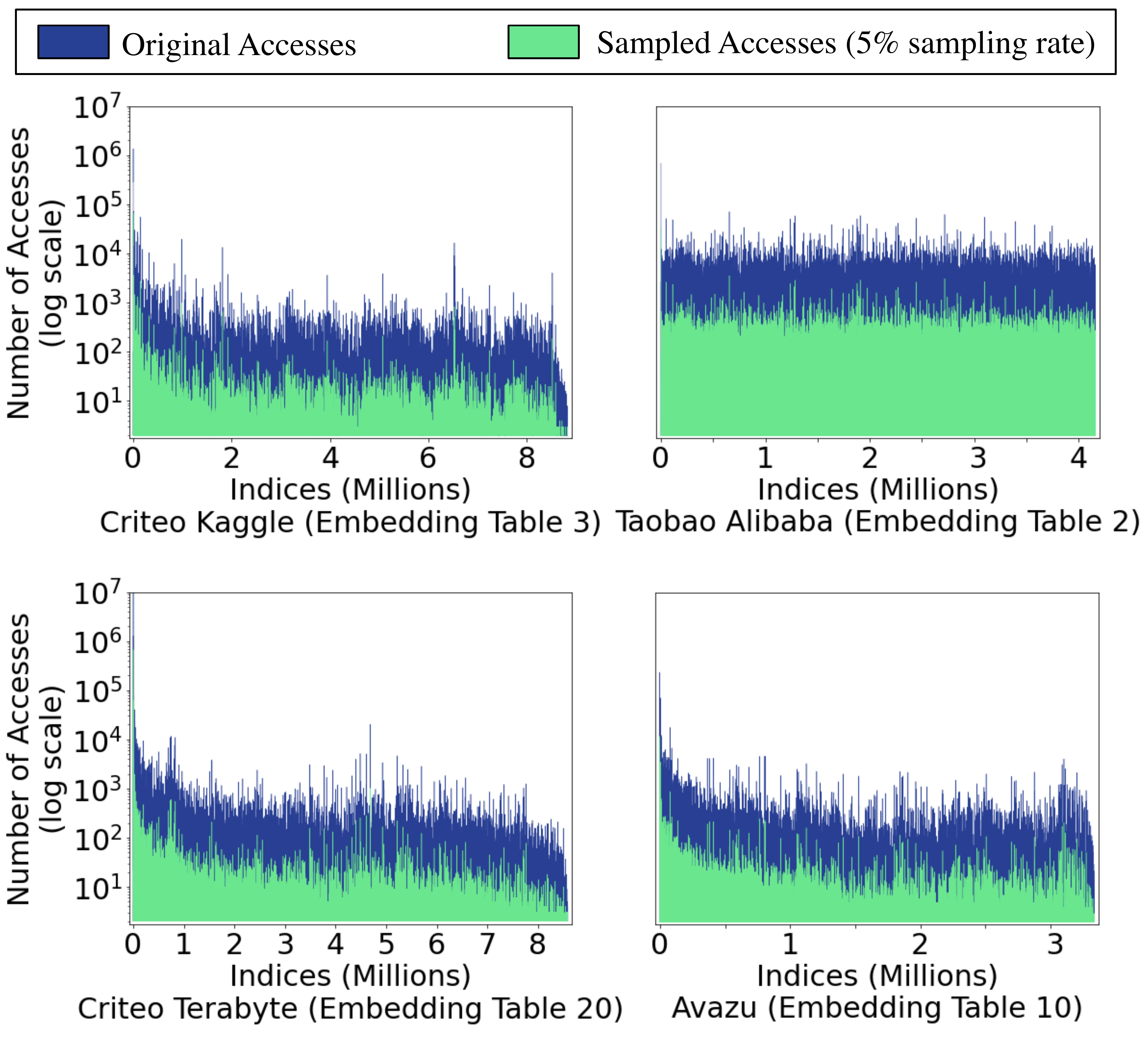}
	\caption{Embedding table access profile from the original inputs (\textbf{\emph{D}}) and the sampled inputs ($\widehat{\textbf{\emph{D}}}$)  -- sampling rate (\textbf{\emph{x}}) = 5\%. We observe that $\widehat{\textbf{\emph{D}}}$ has a similar access signature to \textbf{\emph{D}}.}
	\label{fig:inputsampleaccesspattern}
\end{figure}

As size of the training input dataset is typically very large, we sample \textbf{\emph{x}}\% of the input dataset (\textbf{\emph{D}}). 
The value of \textbf{\emph{x}} is specified as a hyper-parameter. 
Our implementation uses \textbf{\emph{x}} = 5\% and obtains $\widehat{\textbf{\emph{D}}}$ sampled sparse-input entries.
Figure~\ref{fig:inputsampleaccesspattern} shows the access profile for one large embedding table each for Criteo Kaggle, Taobao Alibaba, Criteo Terabyte, and Avazu datasets with and without input sampling.
Empirically, we observe with a sampling rate of 5\%, $\widehat{\textbf{\emph{D}}}$ maintains a similar access signature as \textbf{\emph{D}}.
%
%Using these statistical techniques, the sparse \emph{input sampler} estimates the accesses into embedding tables.
%
%Even with a 5\% sampling rate, we process only $\geq$ 500K inputs.  
%

\begin{figure}[h!]
	\centering
	\includegraphics[width=0.8\columnwidth]{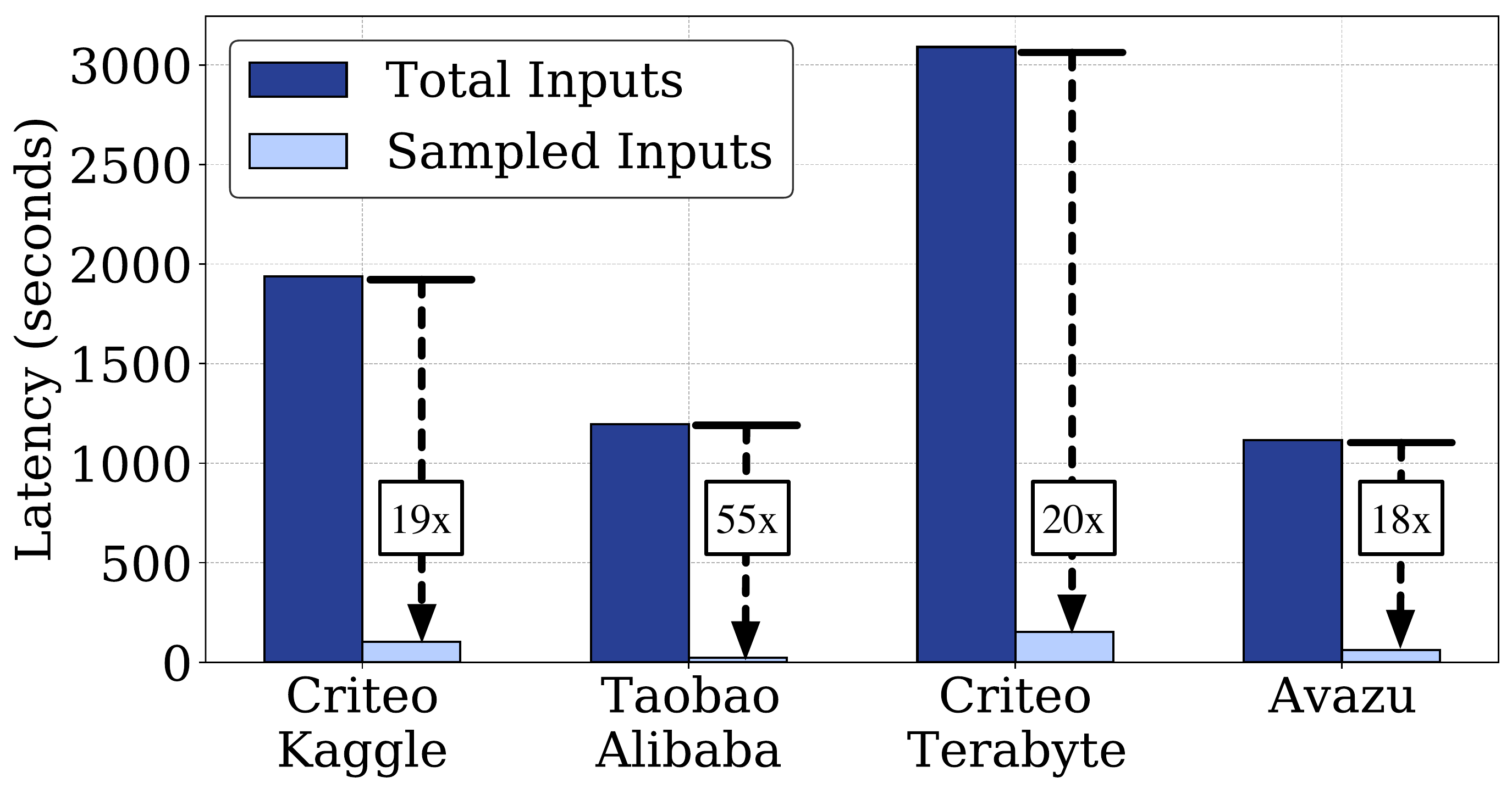}
	\caption{Reduction in the profiling latency when input dataset is sampled for embedding table access pattern. }
	\label{fig:samplelat}
\end{figure}

As shown in Figure~\ref{fig:samplelat}, \framework obtains \sampleinputlatency reduction in latency by input sampling. 
%
%For the Taobao Alibaba dataset, each input has up to 21 sub-inputs enabling a larger drop in latency~\cite{alibaba}. 
%
%Yet, the access signature of frequently accessed entries within embedding tables remains similar.

\subsubsection{Categorize and determine hot embedding size with the Profiler}
\label{sec:logger}
The goal of the \emph{profiler} is twofold - (1) for the sampled input dataset $\widehat{\textbf{\emph{D}}}$ it creates an access profile of each embedding table (\textbf{\emph{E}}$_{z}$), where \textbf{\emph{z}} is the table number and (2) it further samples this access profile to determine what the size of the hot embedding table is.

\niparagraph{Embedding Logger.}
The \emph{profiler} uses an embedding logger for each table to keep track of access counts (denoted as \textbf{\emph{k}}) of $\widehat{\textbf{\emph{D}}}$ into each entry in \textbf{\emph{E}}$_{z}$.
As each model can access multiple embedding tables, \framework assumes any table $\geq$1~MB to be large. 
Embedding tables $<$ 1MB are de-facto considered ``hot'' as they can easily fit even on low-end GPUs.
The \emph{profiler} would still need to estimate the hot embedding table sizes \textit{without} traversing all the embeddings.

\niparagraph{Estimating the hot embedding table sizes per threshold.}
\emph{Profiler} creates a sampled access profile for each embedding table entry across all the tables by selecting random chunks of embedding entries and their observed access pattern from the logger. 
This enables estimating the size of the hot embeddings without traversing all the tables in their entirety.
As the embedding logger observes only \textbf{\emph{x}}\% of the actual inputs, we need to \emph{scale down} the required access counts to classify hot data.
For embedding table number \textbf{\emph{z}} and a threshold \textbf{\emph{t}}, the new hot embedding cutoff for each \emph{sampled entry} is denoted by \textbf{\emph{H}}$_{zt}$, described in Equation~\ref{eqn:cutoff}:

\begin{equation}~\label{eqn:cutoff}
H_{zt} = t \times T \times \frac{x}{100}
\end{equation}
We then pick \textbf{\emph{n}} random samples, each consisting of \textbf{\emph{m}} = 1024 entries entries from embedding logger for table \textbf{\emph{z}}. 
Our implementation uses \textbf{\emph{n}} = 35 and each sample consists of \textbf{\emph{m}} = 1024 embedding entries.
This chunk based sampling allows us to create a distribution of the access pattern.
This paper uses Central Limit Theorem (CLT) to estimate the mean of the parent distribution. 
CLT has the property that, \textit{irrespective of the parent distribution}, the mean of the sampled distribution will always approach the mean of the parent distribution. 
This is because, when the sample size \textbf{\emph{n}} $\geq$ 30, CLT considers the sample size to be large and the sampled mean will be normal even if the sample does not originate from a Normal Distribution~\cite{clt}. 
As each embedding sample chunk consists \textbf{\emph{m}} = 1024 entries, we can estimate the actual embedding table size with a precision of $\frac{1}{1024}$. 
For each chunk, we count (\textbf{\emph{C}}) the number of entries with access counts (\textbf{\emph{k}}) greater than or equal to \textbf{\emph{H}}$_{zt}$. 
%
%The \textbf{\emph{y}} variable is the total number of entries, across all chunks, that have accesses $\geq$ \textbf{\emph{H}}$_{zt}$. 
%
This is represented by Equation~\ref{eqn:chunk}:

\begin{equation}~\label{eqn:chunk}
  C_{i} = \sum_{j=1}^{m} (\text{ $k_{j}$ $\geqslant$ H$_{zt}$}) 
\end{equation}

For \textbf{\emph{n}} chunks, the standard deviation is \textbf{\emph{s}} and the mean is $\bar{\textbf{\emph{y}}}$, shown by Equation~\ref{eqn:samplemean}:
\begin{equation}~\label{eqn:samplemean}
  \bar{y} = \frac{\sum_{i=1}^{n} C_{i}}{n} \\
\end{equation}

\begin{figure}[h!]
	\centering
	\includegraphics[width=0.8\columnwidth]{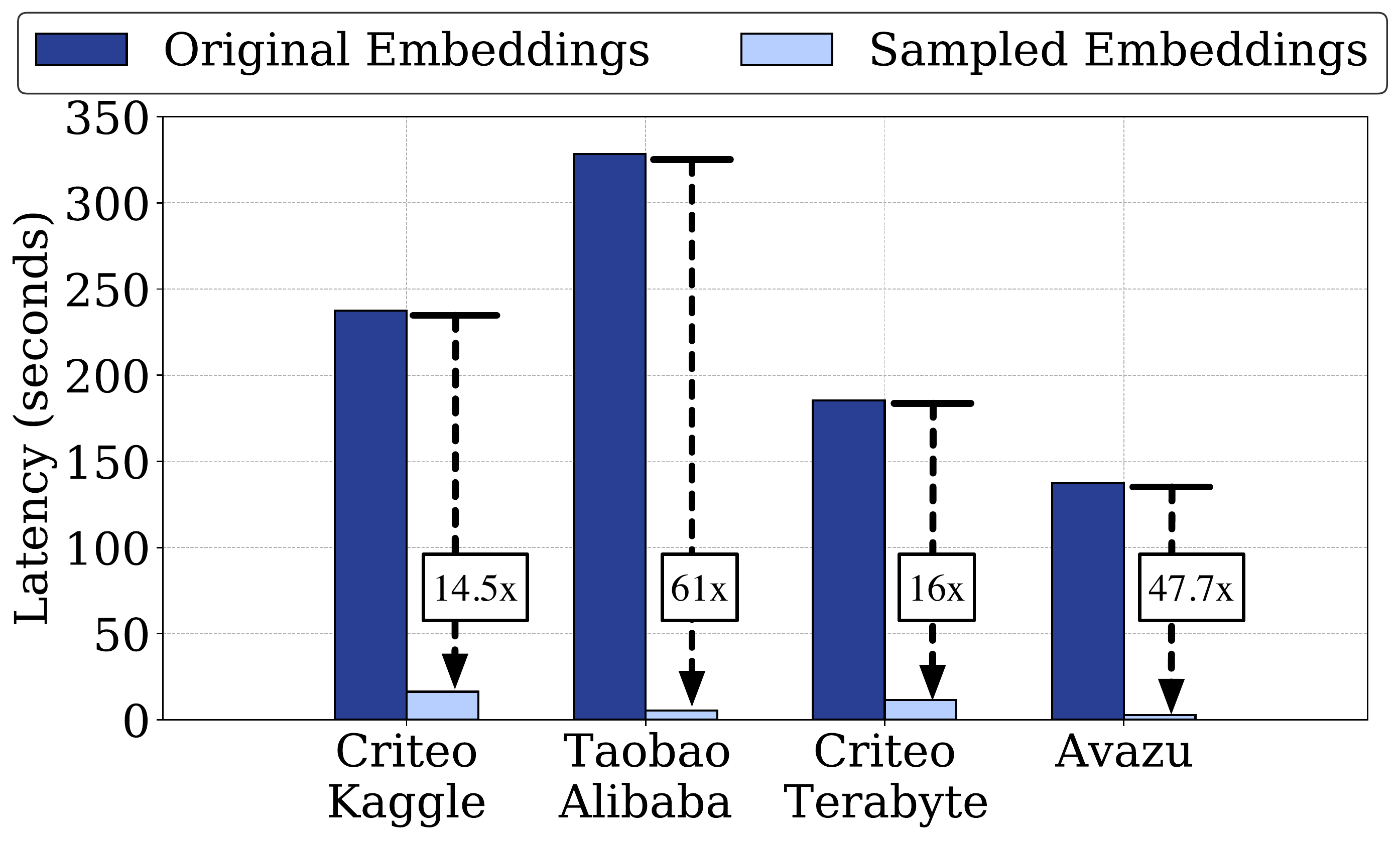}
	\caption{Reduction in the \textit{latency per iteration} by using \embbox to estimate the hot embedding size per threshold. The \textit{total} latency to scan all embedding tables is under 25 seconds per threshold iteration.}
	\label{fig:thresholddelay}
\end{figure}

Figure~\ref{fig:thresholddelay} shows the latency savings from sampling embedding table instead of iterating through all the embedding access content.
%
%Without sampling the embedding table, we would have to iterate throughout the embedding table for each value of $t$.
%
As the \emph{profiler} scans 14x fewer embedding entries for each $t$ it reduces latency of each scan by 14.5$\times$-61$\times$. 

\niparagraph{Input Sampler and Profiler Example:} The Criteo Terabyte dataset is 45~GB in size; post \emph{Input Sampler}, we only process 2.25~GB. 
The \emph{profiler} with this sampled input dataset, logs 8.5M embeddings in the logger for embedding table 20 (\textbf{\emph{E}}$_{20}$),. 
Assuming an interim \textbf{\emph{t}} of $10^{-2}$ and the original training dataset of 60.5M samples, each embedding entry in the logger would have incurred at least 6.05k accesses to be categorized as hot. 
As we use a sampled dataset ($\widehat{\textbf{\emph{D}}}$), the hot entries observe fewer accesses and a smaller threshold of \textbf{\emph{H}}$_{zt}$, 6.05k*$\frac{5}{100}$ = 302.5 accesses.

\niparagraph{Confidence in the estimated embedding table size.}
The goal of the \emph{profiler} is to establish confidence in the estimated embedding size.
A confidence interval, in statistics, refers to the probability ($1-\alpha$) that a population parameter will fall between a set of values.
To compute the confidence interval for the \emph{profiler}'s estimated embedding table size, \framework uses the standard `Student's t-interval'.
As $\bar{\textbf{\emph{y}}}$ follows a t-distribution, the 100$\times$(1-$\alpha$) confidence interval (CI) for $\bar{\textbf{\emph{y}}}$ is represented by Equation~\ref{eqn:confidence}:
\begin{equation}~\label{eqn:confidence}
  CI_{100\times(1-\alpha)} = \bar{y} \pm t_{\frac{\alpha}{2}}\times \sqrt{(\frac{N-n}{N})\times (\frac{s^2}{n})} \\
\end{equation}

\if 0
For a 99.9\% confidence and \textbf{\emph{n}} = 35, the value of $t_{\frac{\alpha}{2}}$ is 3.340. 
%
If \textbf{\emph{N}} is the total number of \textbf{\emph{m}}-sized chucks in each embedding logger, typically \textbf{\emph{N}} $\gg$ \textbf{\emph{n}}.
%
Therefore, we can reduce Equation~\ref{eqn:confidence} to Equation~\ref{eqn:simpleconf}:
\begin{equation}~\label{eqn:simpleconf}
  CI_{99.9} = \bar{y} \pm 3.340 \times \frac{s}{\sqrt{35}} \\
\end{equation}
\fi 

\begin{figure}[h!]
	\centering
	\includegraphics[width=0.9\columnwidth]{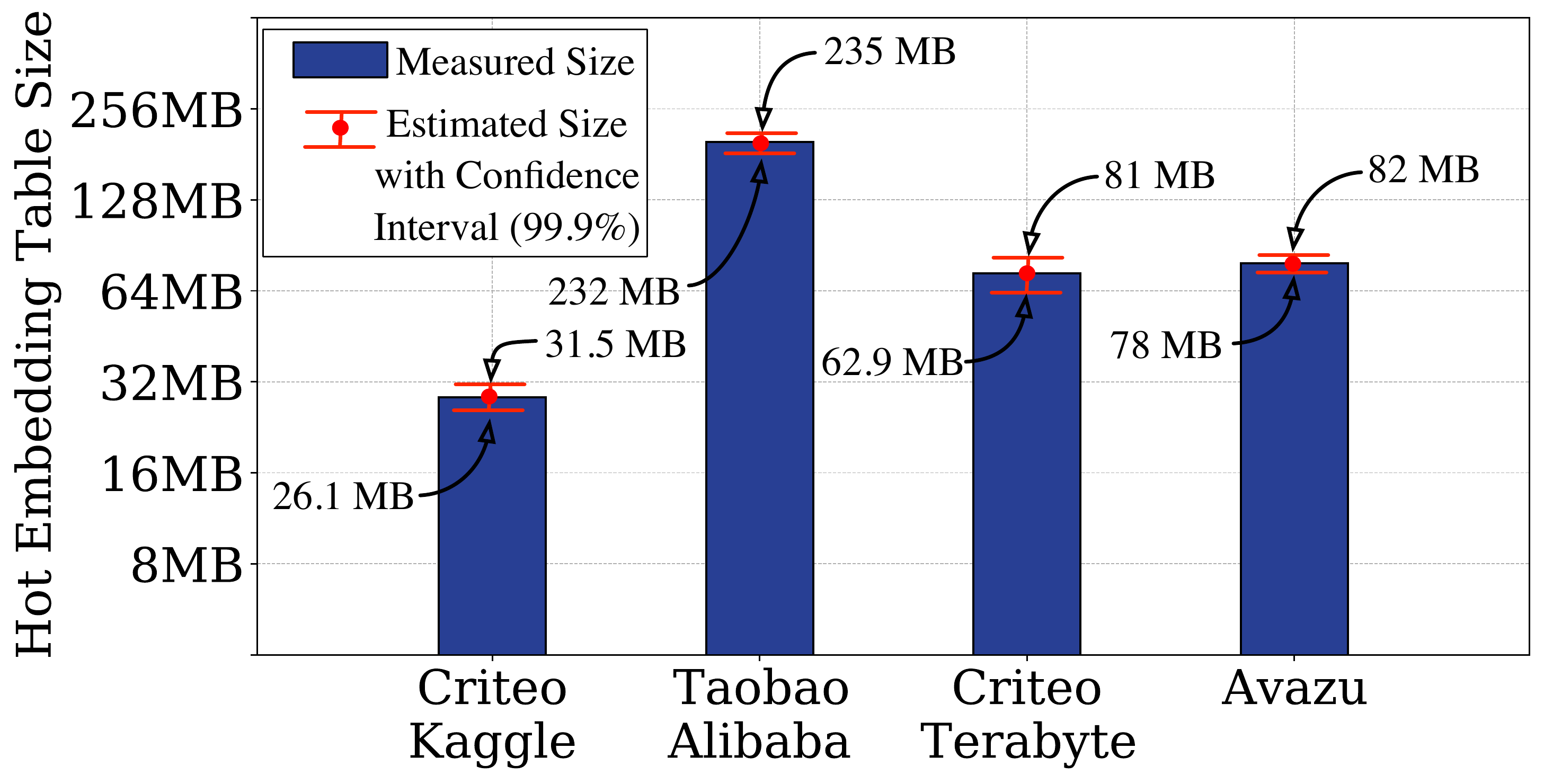}
	\caption{Estimated sizes of hot embedding tables with \embbox. For a confidence interval of 99.9\%, the \embbox estimation is within 10\% (upper bound) of the actual size.}
	\label{fig:embest}
\end{figure}

Figure~\ref{fig:embest} shows the estimation variability compared to the actual values for a confidence interval of 99.9\%.
Actual value of the hot embedding size is the exact size the \emph{profiler} would have obtained if it had processed the entire access pattern for each embedding table. 
This variability can be reduced if we specify a smaller confidence interval.
We observe that the estimated values are within 10\% of the actual values.
As such, for every threshold, the \emph{profiler} process described above is executed to determine the size of the hot embeddings. 
The \statoptimizer, based on this size and user requirements, either accepts the threshold or tunes it further as described below.
Our experiments show that allocated memory of \textbf{\emph{L}} = 512MB suffices for most GPUs (including low-end GPUs).

\subsubsection{Converging on a Threshold using Statistical Optimizer}
The \statoptimizer invokes the profiler with varying \textbf{\emph{t}} (interim thresholds) and a desired confidence interval to determine the final \textbf{\emph{t}}.
Based on the embedding size estimated for an interim threshold, the optimizer tunes the threshold to be higher or lower than the previous one.
This ensures that the threshold is tuned appropriately based on the available GPU memory for each model architecture.
The \statoptimizer then provides the final threshold as output to the next blocks in the \framework. 

\subsection{Input and Embedding Classifier}
\begin{figure}[h!]
	\centering
	\includegraphics[width=0.8\columnwidth]{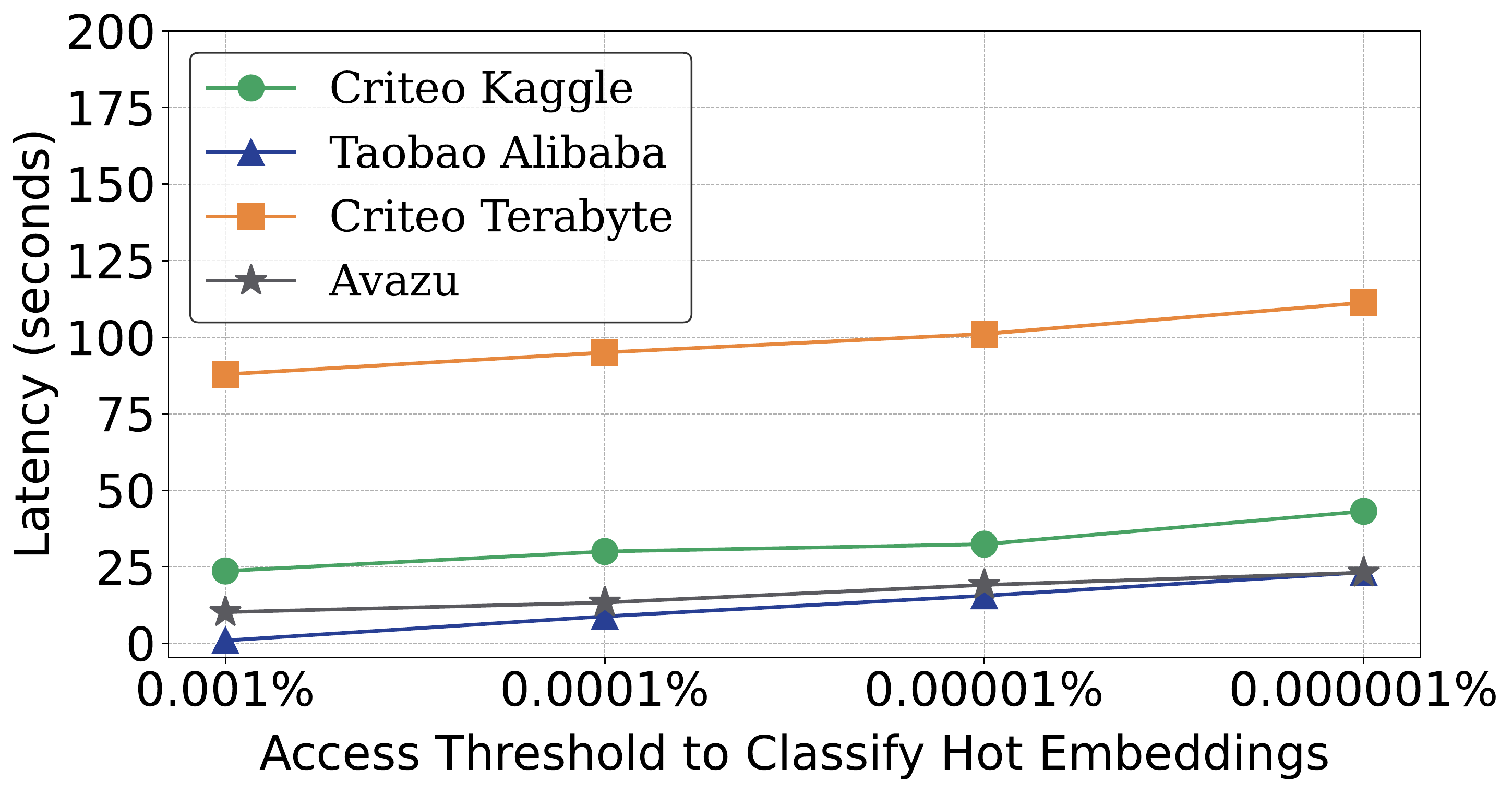}
	\caption{The latency of the input processor to classify sparse-inputs (as hot or cold) as we vary the access threshold. Overall, even for very low access thresholds, we only require only a maximum of 110 seconds.}
	\label{fig:inputprocessor}
	\vspace{-2ex}
\end{figure}

The \emph{embedding classifier} uses the output of the Embedding Logger and the final threshold from \statoptimizer to tag (hot or cold) the embedding table entries. 
This requires \textit{only} one pass of each embedding table. 
%
%During this pass, the Embedding Classifier generates a Hot-Embedding Bag that is sent to each GPU device during the training process.
%
Additionally, the \emph{input classifier} uses the final access threshold value and accesses to the already classified embedding table to identify hot sparse-inputs.
Typically, there are 10s of embedding tables in a recommender model. 
A sparse-input typically accesses one or more entries in each of these embedding tables. 
A sparse-input is classified as hot only if all its embedding table accesses are to hot entries. 
%
%Based on this information, the input processor also forms hot and cold mini-batches (consisting of hot and cold inputs) which can be executed during training.
%
This component typically requires only \textit{one} pass of the entire sparse-input ($S_{I}$) and just checks if the embedding entry indices are present in the hot-embedding bags. 
As this is completely parallelizable operation across both inputs and embedding indices, we divide this task across multiple cores in the CPU.
For a 16 core machine (32 hardware threads), the total time for this phase for different access thresholds is given by Figure~\ref{fig:inputprocessor}.

The \emph{input classifier} also bundles hot and cold inputs together into mini-batches. 
As aforementioned, we require the entire mini-batch to be hot to avoid the data shuffling between CPU and GPU.
If an mini-batch of inputs is entirely hot, the entire execution can happen in a data-parallel mode on the GPU without any interference from the CPU. 
Once we have pre-processed the sparse-input data into hot and cold mini-batches, we store this in the \framework format for any subsequent training runs. 

\subsection{Scheduler for Dynamic Hot-Cold Swaps}
\label{sec:scheduler}

\framework's pre-processing provides a dataset that is distributed into hot and cold mini-batches and a set of hot embeddings.
The \embreplicator replicates the hot embedding bags across all GPUs, but a note here is that the hot embeddings also are available on CPU for baseline cold input executions.
%
%This is because cold inputs can access some hot embedding entries in some tables with simultaneously accessing cold embedding entries in other tables.
%
Next, we discuss the runtime scheduling of hot and cold mini-batches to ensure the baseline accuracy metrics whilst providing accelerated performance.

\begin{table*}
\centering
\caption{Model Architecture Parameters and Characteristics of the Datasets for our Workloads}
\newcolumntype{?}{!{\vrule width 2pt}}
\newcolumntype{P}[1]{>{\centering\arraybackslash}p{#1}}
\setlength\extrarowheight{4pt}
\resizebox{1.9\columnwidth}{!}{
\begin{tabular}{|p{0.14\linewidth}|p{0.16\linewidth}||P{0.08\linewidth}|P{0.06\linewidth}||P{0.04\linewidth}|P{0.04\linewidth}||P{0.06\linewidth}|P{0.09\linewidth}|P{0.06\linewidth}||P{0.14\linewidth}|P{0.11\linewidth}|P{0.09\linewidth}|}
\hline
\multirow{2}{*}{\textbf{Workload}} & \multirow{2}{*}{\textbf{Dataset}} & \multicolumn{2}{c||}{\textbf{Training Input}} & \multicolumn{2}{c||}{\textbf{Model Features}} & \multicolumn{3}{c||}{\textbf{Embedding Tables}} & \multicolumn{3}{c|}{\textbf{Neural Network Configuration}}
\\
\cline{3-12} & & \textbf{Samples} & \textbf{Size} & \textbf{Dense} & \textbf{Sparse} & \textbf{Rows} & \textbf{Row Dim} & \textbf{Size} & \textbf{Bottom MLP} & \textbf{Top MLP} & \textbf{DNN}
\\
\hline
RMC1 (TBSM~\cite{tbsm}) & Taobao (Alibaba)~\cite{alibaba} & 10 M & 1 GB & 3 & 3 & 5.1M & 16 & 0.3 GB & 1-16 \& 22-15-15 & 30-60-1 & Attn. Layer \\
 \hline
RMC2 (DLRM~\cite{dlrm}) & Criteo Kaggle~\cite{criteokaggle} & 45 M & 2.5 GB & 13 & 26 & 33.8M & 16 & 2 GB & 13-512-256-64-16 & 512-256-1 & - \\
 \hline
RMC3 (DLRM~\cite{dlrm}) & Criteo Terabyte~\cite{criteoterabyte} & 80 M  & 45 GB  & 13 & 26  & 266M  & 64 & 63 GB & 13-512-256-64 & 512-512-256-1 & - \\
 \hline
RMC4 (DLRM~\cite{dlrm}) & Avazu~\cite{avazu} & 32.3 M  & 2.4 GB  & 1 & 21  & 9.3M  & 16 & 0.55 GB & 1-512-256-64-16 & 512-256-1 & - \\
 \hline
\end{tabular}}
\label{tab:benchmarks}
\end{table*}

\niparagraph{Impact on accuracy.} 
In the most basic form, \framework can schedule the entire collection of mini-batches comprising hot inputs followed by cold inputs, or vice versa, but such a schedule can have potential impact on training accuracy. 
This is because the hot inputs only access and \emph{update} the hot embedding entries, and training using only hot inputs for a long time can potentially reduce the randomness in training.
For non-convexity loss optimization problems, this makes gradient descent based algorithms susceptible to local minima~\cite{datashuffle, sgdtricks}.
To mitigate this, machine learning community has often deployed data shuffling.
Next, we discuss how we uniquely attenuate this issue for our framework.

\niparagraph{Communication Overheads.} 
To re-introduce randomness in our training while also attaining accelerated performance, we intermittently schedule hot and cold mini-batches.
However, changing input type (hot vs cold) can degrade performance as each of these events requires synchronization of hot embedding parameters between CPU and GPU copies. 
To balance this tradeoff of achieving accuracy but also obtaining performance, we implement \scheduler, a module that dynamically determines the interleaving of hot and cold mini-batches based on the runtime training metric.
The scheduler always begins with training on cold inputs as they update a wider range of embedding entries, albeit infrequently.
The rate of scheduling hot and cold mini-batches can be tuned dynamically based on Equation~\ref{eqn:scheduler}.
In the equation, $r(i)$ is the rate at $i^{th}$ swap. 
Rate of ($R(100)$) implies that 100\% of the mini-batches of cold inputs will be completed before the first hot mini-batches is issued. 
A rate of ($R(1)$) implies hot and cold are shuffled after every mini-batch.
$Test_{L}$ is the testing loss and $u$ is a count of swaps.

\begin{equation}~\label{eqn:scheduler}\small
  r(i+1) =
  \begin{cases}
    min(r(i) * 1/2, R(1)) & if \Delta Test_{L}(i) \geq Test_{L}(i-1) \\
    max(r(i) * 2, R(100)) & if \Delta Test_{L}(i) \leq Test_{L}(i-u) \\
    r(i) & otherwise
  \end{cases}
\end{equation}

Depending on the post-swap testing loss, we change the rate based on two conditions. 
The testing loss used for the scheduler, based on the model requirement, can be loss functions such as mean squared loss and cross-entropy logarithmic loss.
All of our models and their datasets use the logarithmic loss to establish the efficacy of training.
We perform a comparison of loss score between each subsequent swap.
If \framework observes an increase in the test loss, it reduces the rate by half.
This implies that the remaining mini-batches of hot and cold inputs will be split into an alternate of cold and hot schedules. 
The rate can be reduced to a minimum of $R(1)$.

If the test loss decreases, rate remains unchanged, as this is the expected behaviour, unless the loss has been decreasing successively for $u$ schedules.
This is the second case where rate is changed, i.e., increased by 2, up to a max of $R(100)$.
Similar to prior work that offers automatic convergence checks to avoid over-fitting, the downward trend of test loss curve~\cite{earlystopping} consecutively for 4 strips shows a balance between redundancy, badness, and slowness; thus we choose $u$ as 4.
Apart from the above two cases, the rate remains unchanged.
The \scheduler ensures that accuracy remains the priority of \framework. 
\framework begins training with $R(50)$ (alternate cold and hot mini-batches) for a dataset, and tunes the rate accordingly.
%
%Next, we discuss the evaluation of the \framework framework on real world datasets and large recommender models.

%The key insight here is that as inputs are randomly sampled from the entire dataset, the access profile of the $\hat{S_{I}}$ (sampled inputs) will closely track the $S_{I}$ (total dataset).
%
%This insight is bolstered by empirical evidence from real world datasets shown in Figure~\ref{fig:inputsampleaccesspattern}.
%
%Nevertheless, sampling to select only 5\% of the dataset, reduces the latency of accesses per entry by 20x.
%
%\niparagraph{Hot embedding size estimation.}
%
%The \framework framework uses an embedding logger to track the access signatures for $\widehat{\textbf{\emph{D}}}$ inputs. 
%
%This is because, the hot-embedding table sizes depend on the threshold (denoted by $t$) for the number of accesses per entry. 
%
%The \statoptimizer feeds a working value of \textbf{\emph{t}} into the \emph{profiler}. 
%This helps estimate the size of the hot-embedding entries \emph{without} scanning the entire embedding table.
%
%To enable this, the \statoptimizer selects random chunks of embedding entries (samples) from the embedding logger. 
%
%Below, we provide statistical evidence for using a sampler box to determine .
%
\section{Evaluation}
\label{sec:results}

\subsection{Experimental Setup}
\subsubsection{Benchmarks and Real-World Datasets}
\label{sec:benchmarks}

We showcase the efficacy of the \framework framework on 4 real-world datasets, using recommendation models RMC1, RMC2, RMC3, and RMC4. These represent four classes of at-scale models~\cite{facebookreco:hpca}.
We prototype \framework on top of the open source implementation of DLRM~\cite{dlrm} and TBSM~\cite{tbsm}.
There is a model-dataset correspondence, based on the sparse input configuration, with RMC1 model on Taobao Alibaba~\cite{alibaba} with TBSM, and RMC2 on Criteo Kaggle~\cite{criteokaggle}, RMC3 on Criteo Terabyte~\cite{criteoterabyte}, and RMC4 on Avazu~\cite{avazu} with DLRM.
TBSM consists of embedding layer and time series layer (TSL); the embedding layer is implemented through DLRM. TSL resembles an attention mechanism and contains its own MLP network to compute one or more context vectors between history of items and the last item. 
As Taobao Alibaba is the only dataset that provides temporal user-behavior to leverage the TSL layer.
Table~\ref{tab:benchmarks} describes the details of the model architecture for RMC1, RMC2, RMC3, and RMC4 including their dense and sparse features, embedding table numbers and size, and neural network configurations.
In addition to these real world datasets and their corresponding models, we also perform an evaluation on synthetic models. 
As \framework relies on popularity of certain inputs, we execute these synthetic models on Criteo Terabyte (largest dataset) to ensure the semantics of the training input.
%
%Moreover, the table also details the input dataset size used for training.
%
%TBSM architecture is built on DLRM and includes embedding layers, Bottom MLPs, and Top MLPs, like the latter, in addition to a deep learning attention layer.
%
Table~\ref{tab:benchmarks} highlights the diversity of the model architectures in terms of embedding table sizes and neural network configurations.

\begin{figure*}[t]
  \centering
  \subfloat[Criteo Kaggle]{
	\begin{minipage}[t]{0.24\textwidth}
	   \centering
	   \includegraphics[width=\textwidth]{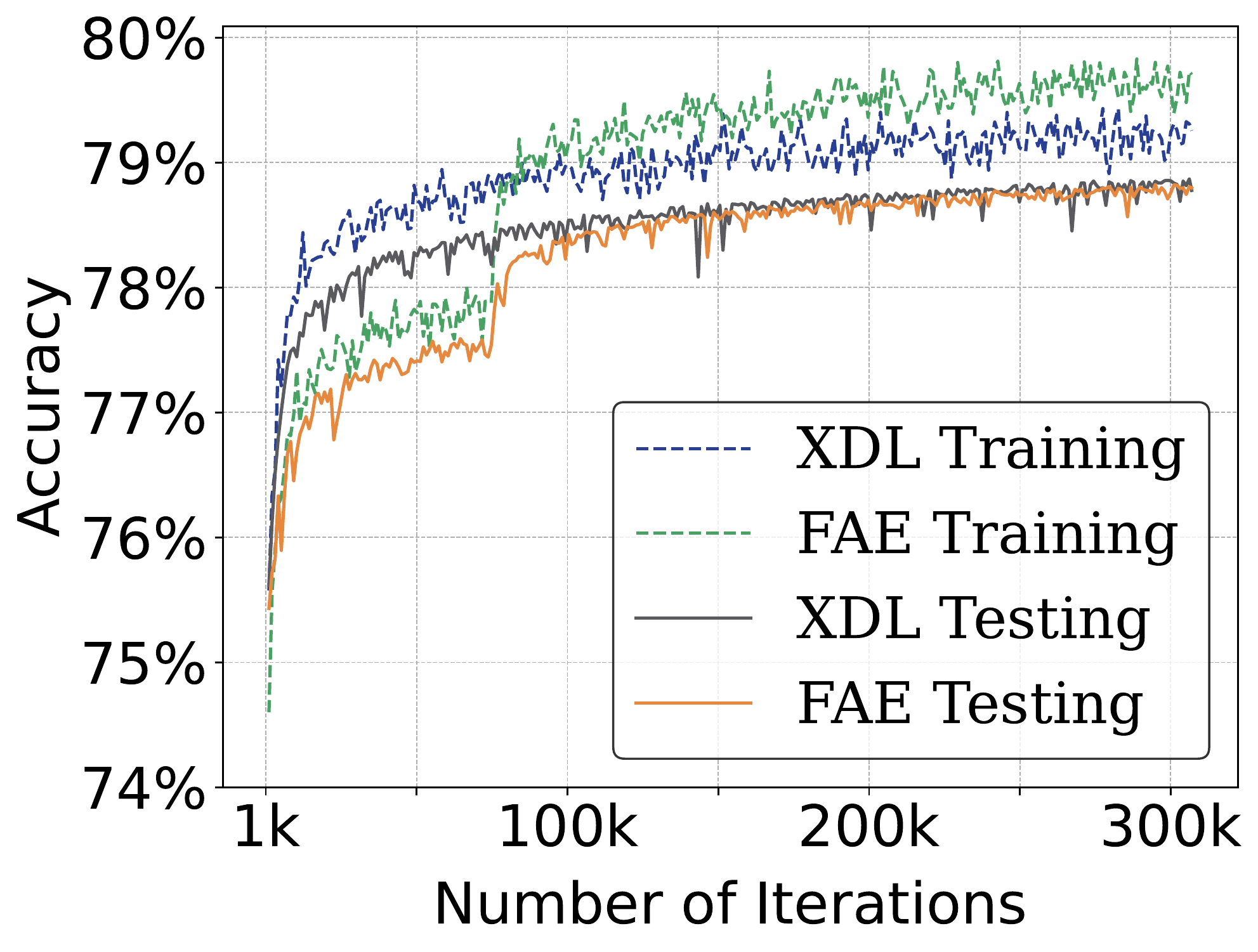}
	\end{minipage}}
 %\hfill 	
  \subfloat[Taobao Alibaba]{
	\begin{minipage}[t]{0.24\textwidth}
	   \centering
	   \includegraphics[width=\textwidth]{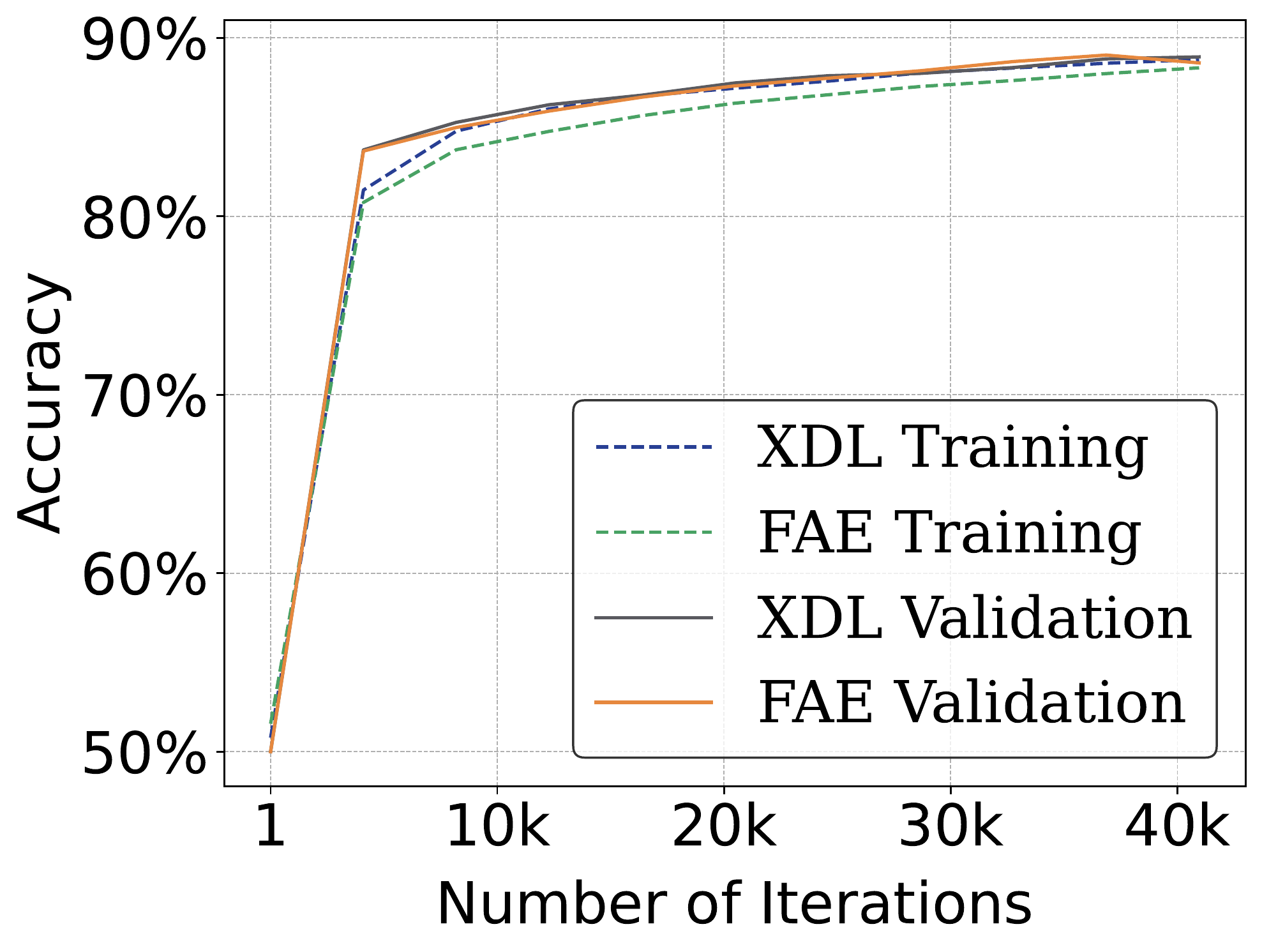}
	\end{minipage}}
 %\hfill	
  \subfloat[Criteo Terabyte]{
	\begin{minipage}[t]{0.24\textwidth}
	   \centering
	   \includegraphics[width=\textwidth]{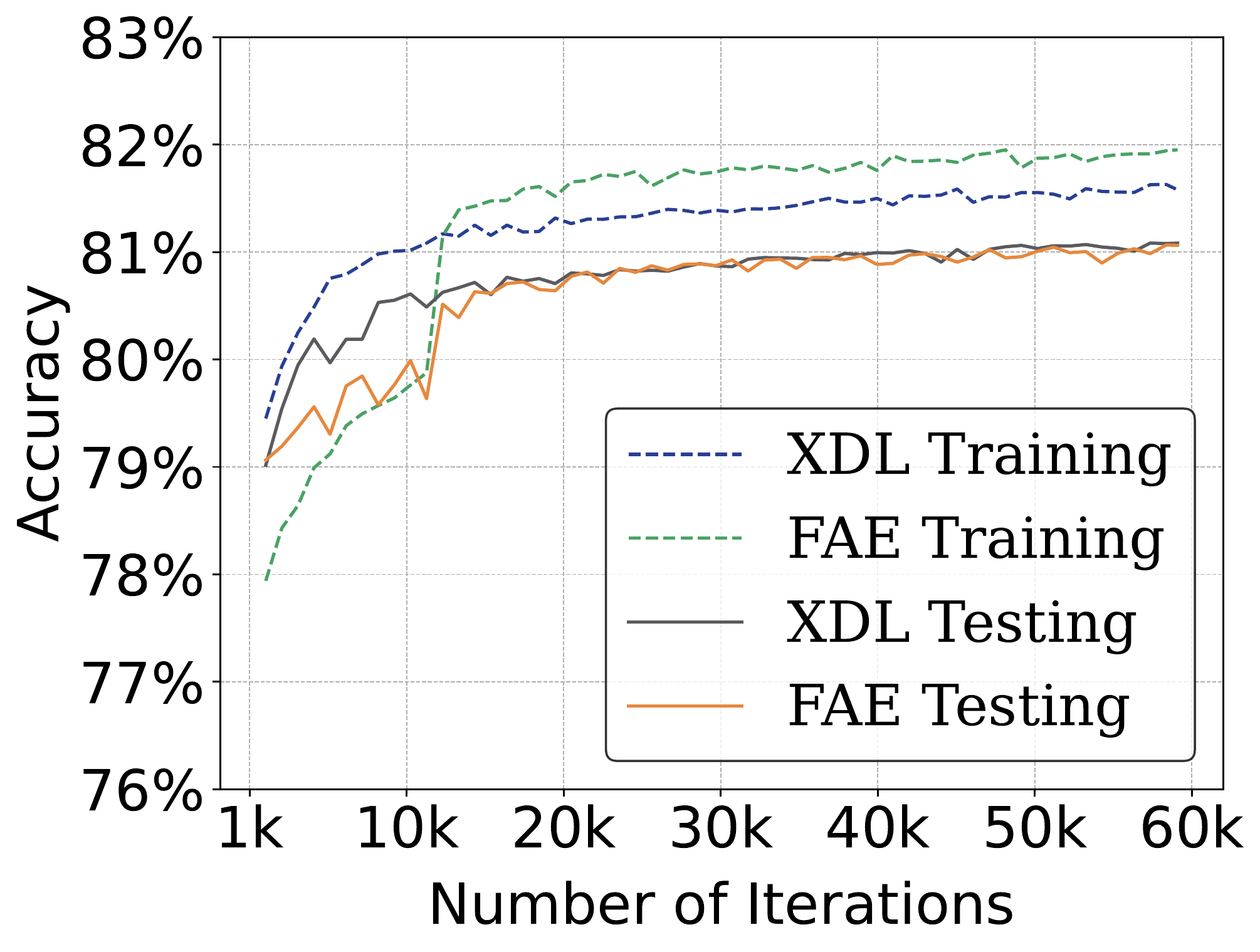}
	\end{minipage}}
 %\hfill	
  \subfloat[Avazu]{
	\begin{minipage}[t]{0.24\textwidth}
	   \centering
	   \includegraphics[width=\textwidth]{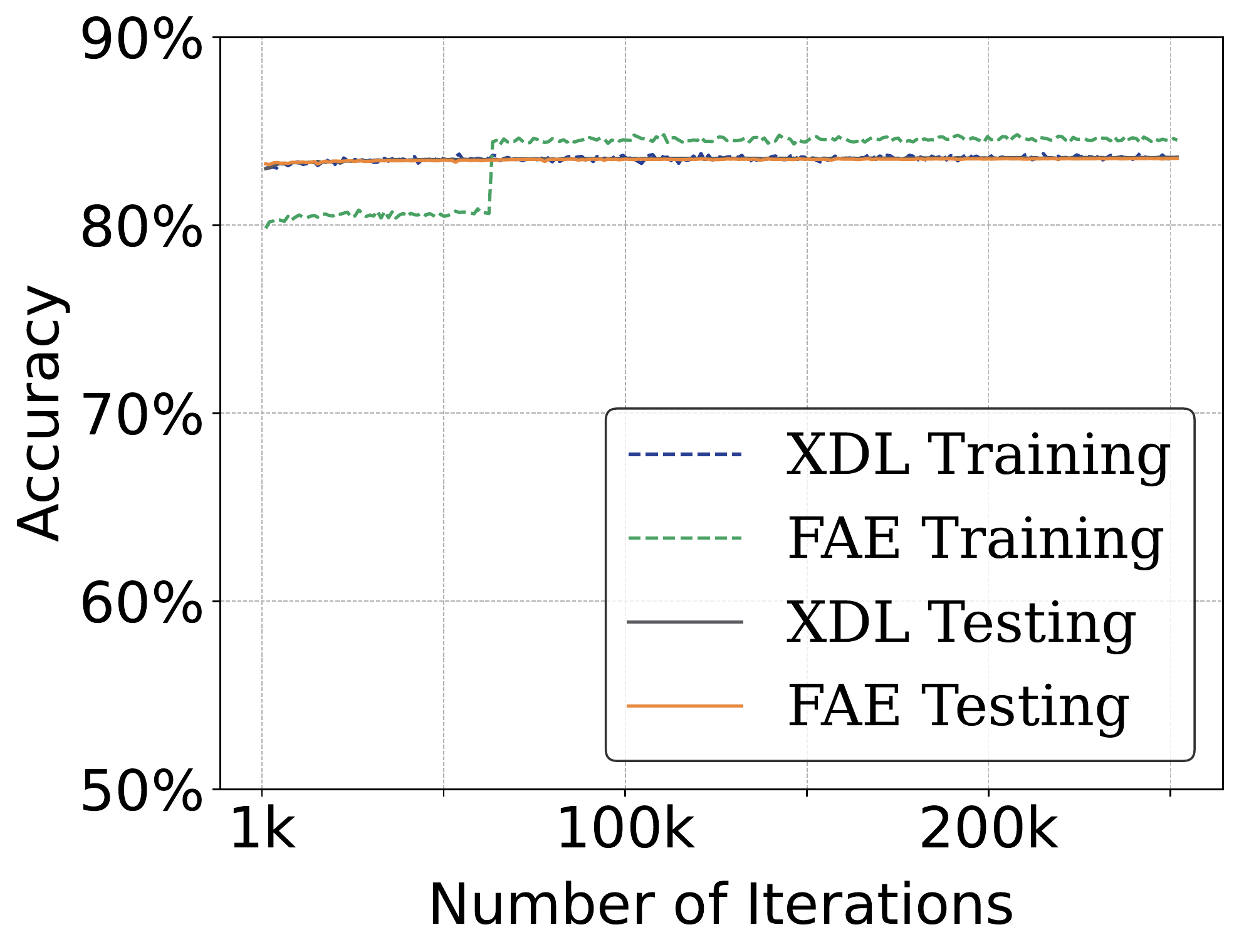}
	\end{minipage}}
\caption{Increasing Accuracy with training iterations when optimized with \framework framework. As we see, all the datasets and corresponding recommender models achieve the XDL accuracy for both training and test or validation sets.}
\label{fig:accuracyresults}
\end{figure*}

\subsubsection{Software libraries and setup}

%In this paper, we use open source DLRM and TBSM recommender systems made available by Facebook Research to evaluate our proposed \framework framework. 
%
The base DLRM and TBSM code is configured using the Pytorch-1.7 and executed using Python-3.
We use the torch.distributed backend to support scalable distributed training and performance optimization~\cite{pytorch}.
NCCL is used~\cite{nccl} for gather, scatter, and all-reduce collective calls via the backend NVLink~\cite{nvlink}.
DLRM and TBSM are also implemented on XDL 1.0~\cite{xdl} using Tensorflow-1.2~\cite{tensorflow} as the computation backend.

\subsubsection{Server Architecture}

Table~\ref{tab:systemconfig} describes the configuration of our datacenter servers \cite{sockeye}.
These servers comprise 24-core Intel Xeon Silver 4116 (2.1 GHz) processor with Skylake architecture.
Each server has a DRAM memory capacity of 192~GB. Each DDR4-2666 channel has 8 GB memory. Each server also has a local storage of 1.9 TB NVMe SSD.
Each server offers 4 NVIDIA Tesla-V100 each with 16GB memory capacity as a general purpose GPU. 
The GPUs are connected using the high speed NVLink-2.0 interconnect.
Every GPU is communicating with the rest of the system via a 16x PCIe Gen3 bus.
In this paper, we perform experiments on a single server with a maximum of 4 GPUs. We expect our insights to hold true even in a multi-server scenario.
\begin{table}[h!]
\centering
\caption{System Specifications}
\resizebox{0.8\columnwidth}{!}{
\begin{tabular}{|c|c|c|c|}
 \hline
 \textbf{Device} & \textbf{Architecture} & \textbf{Memory} & \textbf{Storage}  \\
 \hline \hline
 CPU & Intel Xeon & 768 GB  & 1.9 TB \\
  & Silver 4116 (2.1GHz) & DDR4 (2.7GB/s) & NVMe SSD \\
 \hline
 GPU & Nvidia Tesla & 16 GB  &  - \\
     & V100 (1.2GHz)& HBM-2.0 (900GB/s) &    \\
 \hline

%Then we should add PCIE Here as well? 
\end{tabular}}
\label{tab:systemconfig}
\end{table}

 \if 0
 \hline
 \textbf{Interconnect} & \textbf{System} & \textbf{GPU-GPU} & &
 \hline
 \hline
  & 16x PCIe & NVLink-2.0 & - \\
  & Gen 3 bus & 300 GB/s &  - \\
 \hline
 \fi

\subsubsection{Baselines and terminology.}
We compare \framework optimized training against two baselines: (1) An open source implementation of DLRM and TBSM and (2) A DLRM and TBSM implementation on XDL.
For both baselines we execute on CPU-only mode and CPU-GPU hybrid mode with varying number of GPUs. 
The CPU-only mode is referred to as XDL-CPU and DLRM-CPU.
For CPU-GPU hybrid mode, in case of DLRM, embeddings execute on CPU and neural networks on GPU. For XDL, GPU is used to improve the efficiency of Advance Model Server by using a faster embedding dictionary lookup on GPU. 
CPU is used as a backend worker. 
We represent this mode as X-GPU, where X is the number of GPUs.
\framework optimized training is referred to as X-GPU \framework.

\begin{figure*}[t]
	\centering
	\includegraphics[width=0.95\textwidth]{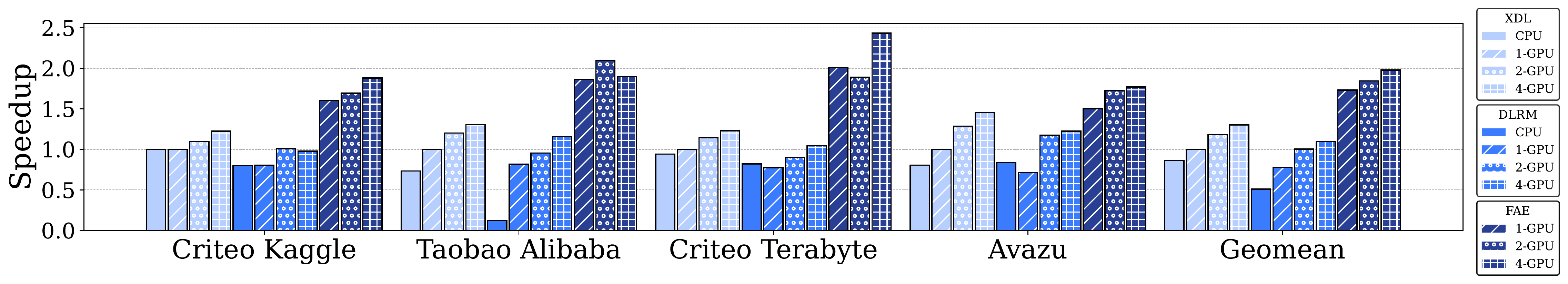}
	\caption{The performance of Criteo Kaggle, Taobao Alibaba, Criteo Terabyte, and Avazu training with the \framework vs XDL and DLRM. All values are normalized to XDL 1-GPU.}
	\label{fig:performance}
\end{figure*}

\begin{figure*}[t]
	\centering
	\includegraphics[width=0.95\textwidth]{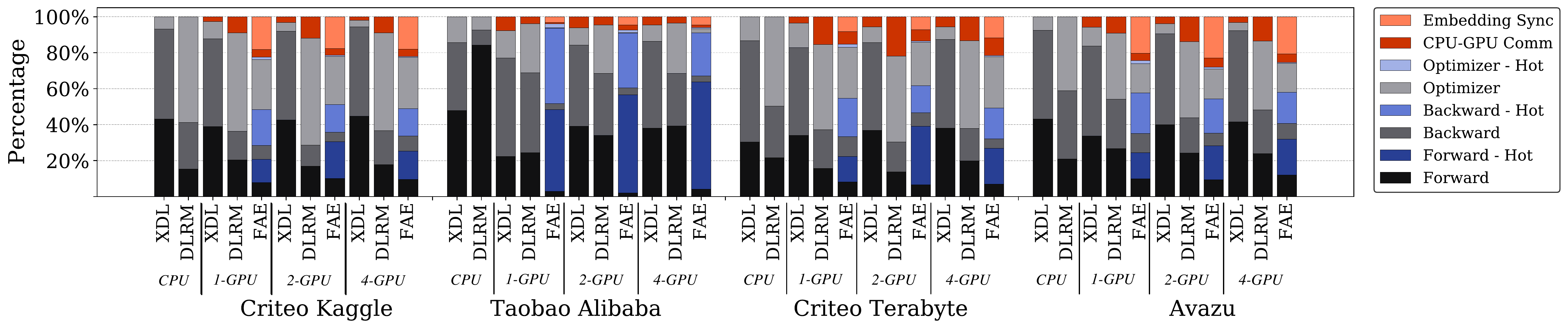}
	\caption{Latency breakdown for the 1, 2, and 4 GPU executions. The \framework framework adds the overhead of embedding synchronization across CPUs and GPUs, not present in XDL and DLRM.}
	\label{fig:latencybreakdown}
\end{figure*}

\subsection{Results and Insights}\label{sec:eval}

\subsubsection{Accuracy Results}
\label{sec:accuracy}
Figure~\ref{fig:accuracyresults} shows the accuracy of Criteo Kaggle, Taobao Alibaba, Criteo Terabyte, and Avazu datasets for their RMC2, RMC1, RMC3, and RMC4 models. 
We use a full-precision XDL-CPU execution baseline. % the baseline remains the same across Section~\ref{sec:eval}.
Table~\ref{table:accuracy} compares the accuracy metrics for all the workloads.
We use testing accuracy, Area Under Curve (AUC), and cross-entropy loss (logloss) as recommendation model performance metric. This metric is established by the MLPerf~\cite{mlperf} community.
For Taobao dataset, we use the accuracy and logloss as performance metric, as AUC is not offered.
As the table shows, each model achieves the corresponding baseline accuracy.
For all the datasets, we observe that when the \scheduler alternately issues cold and hot mini-batches at $R(50)$, the models are able to converge to the baseline accuracy in the same number of baseline training iterations. 
%
%For the Criteo Terabyte dataset even a cold followed by hot mini-batch scheduling suffices and the optimized training achieves the baseline accuracy. 
%
\framework observes an initial jump in accuracy for both Criteo and Avazu datasets after the first swap between cold and hot mini-batch. 
Once, the model is trained on both the types of mini-batches, we do not observe any more jumps. 
As we interleave it with the first hot mini-batch, many pertinent embedding entries get updated and we reach the baseline accuracy for both training and testing sets.
\begin{table}[h!]
\centering
\caption{Accuracy Metric Comparisons}
\scriptsize
\newcolumntype{?}{!{\vrule width 2pt}}
\newcolumntype{P}[1]{>{\centering\arraybackslash}p{#1}}
\setlength\extrarowheight{4pt}
\resizebox{\columnwidth}{!}{
\begin{tabular}{p{0.16\linewidth}|P{0.15\linewidth}P{0.05\linewidth}P{0.07\linewidth}|P{0.15\linewidth}P{0.05\linewidth}P{0.07\linewidth}}
\hline
\multirow{2}{*}{\textbf{Dataset}} &  \multicolumn{3}{c}{\textbf{XDL}} & \multicolumn{3}{|c}{\textbf{FAE}}
\\
\cline{2-7} & \textbf{Accuracy (\%)} & \textbf{AUC} & \textbf{Logloss} & \textbf{Accuracy (\%)} & \textbf{AUC} & \textbf{Logloss}
\\
\hline
\hline
Criteo Kaggle & 78.86 & 0.802 & 0.452 & 78.86 & 0.802 & 0.452 \\
 %\hline
 Taobao Alibaba & 89.21 & - & 0.269 & 89.03 & - & 0.271 \\
 %\hline
 Criteo Terabyte & 81.07 & 0.802 & 0.424 & 81.06 & 0.802 & 0.424 \\
 %\hline
 Avazu & 83.61 & 0.758 & 0.390 & 83.60 & 0.758 & 0.391 \\
 \hline

\end{tabular}}
\label{table:accuracy}
\end{table}

%
%For training set \framework framework can even surpass the baseline accuracy from to xxx to xxx and to xxx to xxx for criteo kaggle and criteo terabyte.

%shows the overall result accuracy comparing baseline and FAE. Using figure 15a, we can notice that FAE training accuracy surpasses baseline training accuracy in the end. In the beginning of training we start with processing normal data that access both hot and not hot embedding table entries. Depends on minibatch size at some point during the training we start switching to hot data and the shoot up point is the result of that. After switching from feeding normal input to hot input, we can see that FAE reaches a higher accuracy than the baseline during training phase.
%
%During validation, using normal data in the beginning, test accuracy of FAE is less than baseline. That's because we use normal data to train bigger portion of embedding tables as they access both hot and not hot area of embedding tables. That being said, when we switch to testing with hot inputs, we already semi-trained significant portion of embedding tables compared to otherwise.

\subsubsection{Performance Gains and Absolute Training Times}
\label{sec:evalperformance}
Figure~\ref{fig:performance} shows the performance improvement of end-to-end training execution using \framework in comparison to XDL and DLRM/TBSM. 
The end-to-end training runs are terminated when the established accuracy metric (cross-entropy loss or area under the curve) is met. 
The performance is normalized to XDL 1-GPU execution
%The performance of CPU-only, hybrid, and FAE modes are normalized to a 1-GPU hybrid for each dataset.
%
%In the figure the baseline is consistently 1 GPU baseline execution of each dataset.
%
For a single device (CPU or 1-GPU), we use a mini-batch of \kagglembs, \taobaombs, \terabytembs and \avazumbs inputs for Criteo Kaggle, Taobao Alibaba, Criteo Terabyte and Avazu, respectively.
\framework training reduces the average execution time (geomean) by 42\%, 36\%, and 34\%, 1-GPU, 2-GPU, and 4-GPU executions, respectively.
The GPU comparisons assume same number of GPUs for XDL and \framework.
We maintain weak scaling across distributed runs where the mini-batch size is scaled with the number of GPUs.
For example, 2 GPU execution use 2K, 512, 2K and 2K mini-batch size for Criteo Kaggle, Taobao Alibaba, Criteo Terabyte and Avazu, respectively.
For Taobao, 4 GPU execution takes more time than 2 GPU execution because the dataset is relatively small, thus the cold mini-batch executions overshadow benefits of \framework.
Overall \framework reduces the training time by \improvementavgxdlcpu and \improvementavgxdlgpu in comparison to XDL CPU-only and XDL CPU-GPU with 4-GPUs.

\begin{table}[h!]
\centering
\caption{Absolute Training Time for 10 Epochs (mins)}
\scriptsize
\newcolumntype{?}{!{\vrule width 2pt}}
\newcolumntype{P}[1]{>{\centering\arraybackslash}p{#1}}
\setlength\extrarowheight{4pt}
\resizebox{\columnwidth}{!}{
\begin{tabular}{|p{0.158\linewidth}|P{0.08\linewidth}|P{0.064\linewidth}|P{0.043\linewidth}|P{0.064\linewidth}|P{0.043\linewidth}|P{0.064\linewidth}|P{0.043\linewidth}|}
\hline
\multirow{2}{*}{\textbf{Dataset}} & 
\multirow{2}{*}{\textbf{XDL}} &
\multicolumn{2}{c|}{\textbf{1-GPU}} & \multicolumn{2}{c|}{\textbf{2-GPU}} & \multicolumn{2}{c|}{\textbf{4-GPU}}
\\
\cline{3-8} & \textbf{CPU} & \textbf{XDL} & \textbf{FAE} & \textbf{XDL} & \textbf{FAE} & \textbf{XDL} & \textbf{FAE}
\\
\hline
 Criteo Kaggle & 197.56 & 196.97 & 122.71 & 179.16 & 116.27 & 160.65 & 104.69 \\
 \hline
 Taobao Alibaba & 1108.84 & 813.10 & 436.58 & 677.00 & 387.79 & 621.96 & 428.55 \\
 \hline
 Criteo Terabyte & 404.25 & 380.88 & 189.73 & 330.06 & 201.61 & 309.51 & 156.45 \\
 \hline
 Avazu & 134.28 & 108.24 & 72.07 & 84.04 & 62.73 & 74.20 & 61.15 \\
 \hline

\end{tabular}}
\label{table:abs_trainingtime}
\vspace{-2ex}
\end{table}

\niparagraph{Absolute time:} 
Table~\ref{table:abs_trainingtime} shows the absolute end-to-end training time, when the executions reach their required accuracy metric.
We use minibatch of 1k, 2k, and 4k for Crtieo Kaggle, Terabyte, and Avazu datasets. We use minibatch of 256, 512 and 1k for the Taobao Alibaba dataset.
We observe that the RMC1 model with Taobao Alibaba dataset obtains most benefits from GPU acceleration as it employs a relatively large DNN.
\framework can further accelerate the training of this model and reduce the training time to 428 minutes with 4-GPU \framework compared to 621 minutes with 4-GPU XDL.
%
%With the recent developments in machine learning~\cite{attention}, we expect the neural networks in recommender models to considerably increase in size. 
%
Our results clearly show that \framework can enable GPU acceleration without incurring large data CPU-GPU transfer overheads.  

\subsubsection{Latency breakdown.}
Figure~\ref{fig:latencybreakdown} shows the breakdown of the total runtime for each of the workloads executing on CPU-only and 1, 2, and 4 GPUs.
%
%We employ a weak scaling technique where 1-GPU execution uses \kagglembs, \taobaombs, \terabytembs and \avazumbs  mini-batch size for Criteo Kaggle, Taobao Alibaba, Criteo Terabyte and Avazu datasets, respectively.
%
%These results are also when the mini-batch size is scaled as the number of devices increase.
%
The breakdown for cold inputs are consistent across XDL, DLRM, and \framework executions. 
%
%We observe that the optimizer constitutes a large portion of the DLRM execution.
%
%This is because, a CPU cannot efficiently execute the massively parallel optimizer operation.
%
\framework is able to mitigate some of these inefficiencies by performing both the neural network and embedding updates on GPUs for the hot input mini-batches.
In case of XDL, efficiency of Advanced Model Server (AMS) is improved using GPU to speed up the massively parallel optimizer and embedding dictionary lookup. 
Even XDL is limited by the size of GPU memory, hence only the index of embedding dictionary is stored in GPU memory.
%
%The optimizer, such as Stochastic Gradient Descent, is massively parallel and therefore is not well suited for CPU execution.
%
Due to small size of hot embedding tables, \framework stores the entire table in GPU memory instead of only the indices.
%
%Hence, for \framework the optimizer time for hot mini-batches is significantly lower than for the cold mini-batches, as the of hot inputs are observed more often but accelerated on GPU.
%
Figure~\ref{fig:latencybreakdown} also shows the percentage of time spent by XDL, DLRM/TBSM, and \framework on embedding layer data transfer.
This data transfer is completely eliminated for \framework for hot mini-batches. 
For DLRM/TBSM implementations, the data transfer time comprises the time spent on transferring embedding data to the GPU. For XDL, the time reported is spent on transferring embedding indices and model dense parameters to the GPU.

\niparagraph{Embedding Synchronization:} One overhead imposed by \framework is from embedding synchronization while switching between cold and hot mini-batches. 
The embedding tables are updated across CPU and GPU memories to ensure the training process observes the same entries. 
This overhead is shown by the `embedding sync' entry Figure~\ref{fig:latencybreakdown}.
The Avazu dataset observes a higher percentage of embedding synchronization overhead because of its comparatively smaller embedding size.
Thus the fixed transfer cost from CPU to GPU, using PCIe, is not amortized over a large data transfer.
On the contrary, the Taobao dataset observes the least percentage of synchronization overhead.
This can be attributed to the high percent of forward and backward time of the RMC1 recommender model due to its deep attention layer.
Thus, as the recommender models become bigger with larger embedding tables and deeper neural network layers, \framework can offer higher benefits by reducing the CPU-GPU data transfer between embedding and DNN layers, whilst observing amortized embedding synchronization overheads.
%
%This is because, even though embedding tables are expected to increase in size, a larger absolute size of embeddinga does not necessarily imply a proportionally large hot embedding table.
%
%This is because certain inputs are always going to be way more popular than the others.

%Proposed framework tends to segregate inputs that are accessing only hot rows named hot inputs from not hot or cold input that access both hot or cold data. If we process all embedding requests on GPU, cold inputs incur a significant communication overhead between GPUs. By employing FAE, we notice hot regions across all embedding table.We expect to the most gain of speed up for larger datasets as they tend to contain more cold data and that is equivalent to more potential communication overhead.
%Based on figure 12, we can see that the most speed up is for RMC3 using Criteo Terabyte with overall size of 61GB and 80M inputs.
%Using Geomean, on average, We achieved more than 2x speedup using FAE in comparison to baseline.

\begin{table}[h!]
\centering
\caption{CPU-GPU data transfer time for 10 Epochs (mins)}
\scriptsize
\newcolumntype{?}{!{\vrule width 2pt}}
\setlength\extrarowheight{4pt}
\resizebox{\columnwidth}{!}{
\begin{tabular}{|p{0.16\linewidth}|p{0.06\linewidth}|p{0.06\linewidth}|p{0.06\linewidth}|p{0.06\linewidth}|p{0.06\linewidth}|p{0.06\linewidth}|p{0.06\linewidth}|p{0.06\linewidth}|p{0.06\linewidth}|}
\hline
\multirow{2}{*}{\textbf{Dataset}} &  \multicolumn{3}{c|}{\textbf{1-GPU}} & \multicolumn{3}{c|}{\textbf{2-GPU}} & \multicolumn{3}{c|}{\textbf{4-GPU}}
\\
\cline{2-10} & \textbf{DLRM} & \textbf{XDL} & \textbf{FAE} & \textbf{DLRM} & \textbf{XDL} & \textbf{FAE} & \textbf{DLRM} & \textbf{XDL} & \textbf{FAE}
\\
\hline
 Criteo Kaggle & 22.09 & 5.39 & 4.99 & 23.12 & 5.61 & 4.35 & 18.00 & 3.05 & 4.29 \\
 \hline
 Taobao Alibaba & 37.93 & 24.97 & 3.24 & 38.27 & 12.89 & 11.11 & 25.04 & 6.24 & 6.04 \\
 \hline
 Criteo Terabyte & 76.01 & 13.46 & 13.27 & 92.98 & 18.94 & 12.41 & 48.43 & 17.49 & 15.24 \\
 \hline
 Avazu & 13.94 & 6.23 & 2.97 & 12.68 & 3.19 & 3.17 & 11.94 & 2.36 & 2.79 \\
 \hline

\end{tabular}}
\label{table:cpu_gpu_time}
\end{table}

\niparagraph{Data transfer between CPU and GPU.}
Table~\ref{table:cpu_gpu_time} shows the absolute communication time to transfer the embedding layers and Table~\ref{table:cpu_gpu_data} shows the amount of data transferred for XDL, DLRM/TBSM and \framework execution including the embedding synchronization for \framework.
On average \framework reduces the total data transfer from 37 GB with XDL to 24 GB, even including the embedding synchronization overhead, which translates to 12\% improvement in CPU-GPU data transfer time.
In case of XDL, all dense parameters needs to be transferred from AMS to backend workers and vice versa per training iteration.
\framework only require parameters to be transferred between CPU and GPU across the hot and cold mini-batch swap.
%
%While for \framework all dense parameters are only replicated along with hot embeddings across all GPU's using high speed NVLink interconnects.}

\begin{table}[h!]
\centering
\caption{Amount of Data Transferred over 10 Epochs}
\resizebox{0.8\columnwidth}{!}{
\begin{tabular}{|c|c|c|c|}
 \hline
 \textbf{Dataset} & \textbf{DLRM (GB)} & \textbf{XDL (GB)} & \textbf{FAE (GB)}  \\
 \hline \hline
 Criteo Kaggle & 60.89 & 23.16 & 14.99 \\
 \hline
 Taobao Alibaba & 1.95 & 0.51 & 0.61 \\
 \hline
 Criteo Terabyte & 375.06 & 95.60 & 69.58\\
 \hline
 Avazu & 40.45 & 30.27 & 10.45 \\
 \hline
\end{tabular}}
\label{table:cpu_gpu_data}
\end{table}

\subsubsection{Performance improvement with varying mini-batch size}
\begin{figure}[h!]
	\centering
	\includegraphics[width=1\columnwidth]{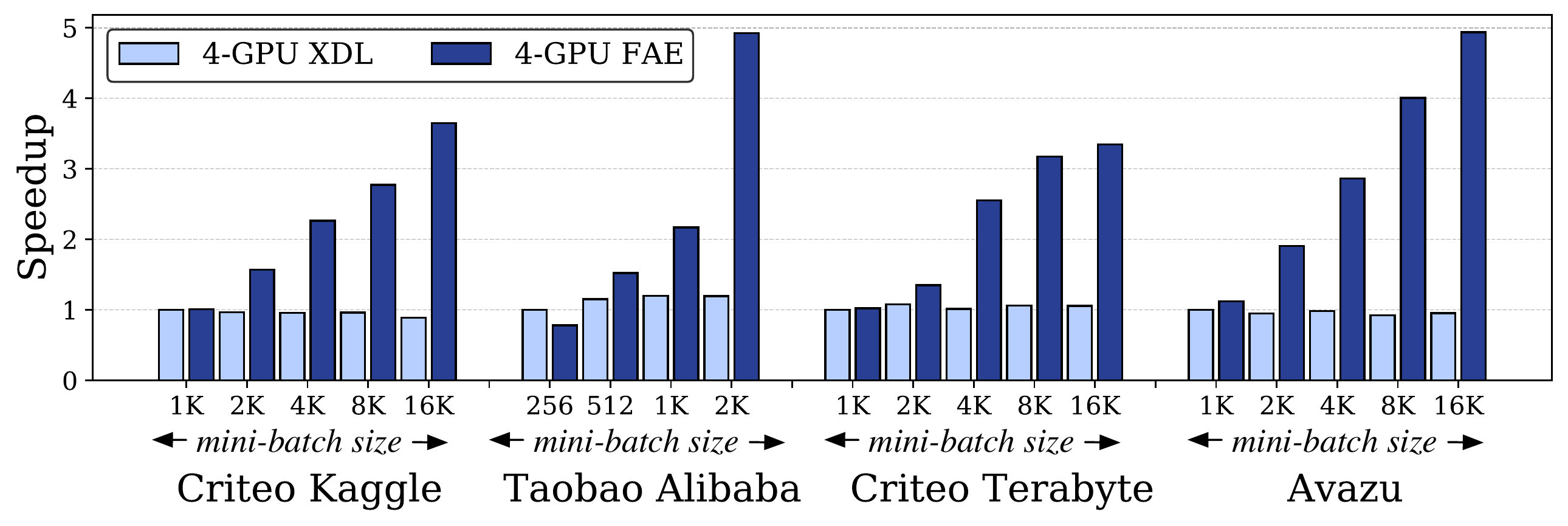}
	\caption{Speedup of \framework with varying mini-batch sizes for a 4-GPU system, compared to a 4-GPU XDL}.
	\label{fig:minibatch_perf}
\end{figure}

Figure~\ref{fig:minibatch_perf} shows the performance benefits of \framework training over XDL execution for a 4-GPU system.
Speedup is normalized to XDL execution with mini-batch size of \kagglembs, \taobaombs, \terabytembs and \avazumbs~for Criteo Kaggle, Taobao Alibaba, Criteo Terabyte and Avazu datasets respectively. 
As the mini-batch size increases, we observe higher benefits because the overheads of \framework are amortized over a larger input set.
For instance, now the \embreplicator replicates the model fewer times.
However, with XDL, we do not see such an improvement because of extra time being spent on creating and sending larger mini-batches to the backend workers.

\subsubsection{Performance improvement for synthetic models}
\label{sec:synthetic}
\begin{figure}[h!]
	\centering
	\includegraphics[width=0.48\textwidth]{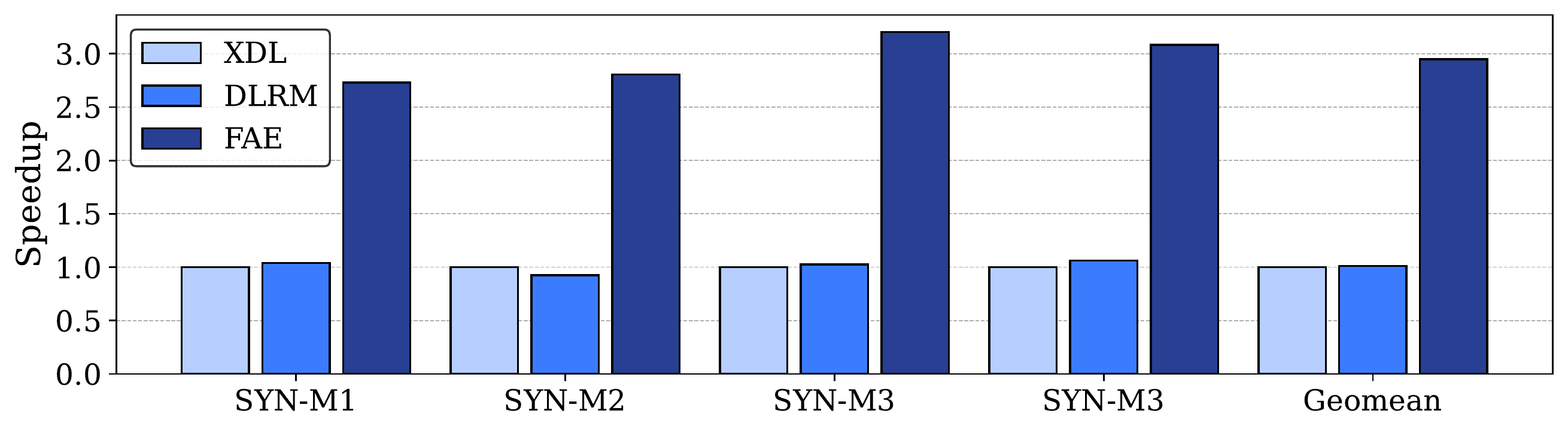}
	\caption{Performance comparison of \framework with XDL 4-GPU across various synthetic models.}
	\label{fig:synthetic_perf}
\end{figure}

%We use real-world training data as \framework utilizes the property that certain inputs are way more popular than the others. 
%
To understand the efficacy of \framework on varying types of model architectures, we create synthetic configurations, shown in Table~\ref{table:syntheticmodels},  to execute the Terabyte dataset.
Figure~\ref{fig:synthetic_perf} shows the speedup of \framework across various synthetic models. 
\framework provides 2.94$\times$ average speedup across small and large synthetic models as compared to XDL.
%
%In future, if some other large models were utilized instead of MLP, we expect \framework to perform well for those models also as we execute DNN's on GPU as compared to backend workers for XDL which are CPU centric most of the times.}
\begin{table}[h!]
\centering
\caption{Synthetic Models' Configuration}
\vspace{-0.1in}
\resizebox{0.7\columnwidth}{!}{
\begin{tabular}{|c|c|c|}
 %\multicolumn{4}{|c|}{Power Specification} \\
 \hline
 \textbf{Dataset}& \textbf{Bottom MLP} & \textbf{Top MLP} \\
 \hline
 \hline
 SYN-M1 & 13-64 & 512-1 \\
  \hline
 SYN-M2 & 13-512-64 & 512-256-1 \\
  \hline
 SYN-M3 & 13-1024-512-64 & 512-1024-256-1 \\
  \hline
 SYN-M4 & 13-1024-512-256-64 & 512-1024-512-256-1 \\
  \hline
\end{tabular}}
\label{table:syntheticmodels}
\vspace{-2ex}
\end{table}

\subsubsection{Power Benefits}
Table~\ref{table:powernumbers} shows the per GPU power consumption using the baseline and \framework for a 1024 mini-batch. 
\framework reduces GPU power consumption by 9.7\% in comparison to XDL.
This is primarily due to the reduced communication cost between devices.
\begin{table}[h!]
\centering
\caption{GPU Power Consumption Comparison}
\resizebox{0.65\columnwidth}{!}{
\begin{tabular}{|c|c|c|c|}
 %\multicolumn{4}{|c|}{Power Specification} \\
 \hline
 \textbf{Dataset}& \textbf{XDL} & \textbf{DLRM} & \textbf{FAE}  \\
 \hline
 \hline
 Criteo Kaggle & 61.83W & 58.91W & 55.81W \\
  \hline
 Alibaba & 56.39W & 60.21W & 56.62W \\
 \hline
 Criteo Terabyte & 59.71W & 62.47W & 57.03W \\
 \hline
 Avazu & 60.2W & 58.03W & 56.4W \\
 \hline
\end{tabular}}
\label{table:powernumbers}
\end{table}

\if 0
 %Criteo Kaggle & 2048 & 58.53W & 57.55W \\
 \hline
 %Criteo Kaggle & 4096 & 60.36W & 56.91W \\
\hline
 %Criteo Kaggle & 8192 & 58.72W & 54.44W \\
 \hline
 %Criteo Kaggle & 16384 & 59.04W & 56.19W \\
 \hline
 %Alibaba & 256 & 60.21W	& 56.62W \\
 \hline
 %Alibaba & 512 & 60.21W	& 56.62W \\
 \hline
 Alibaba & 1024 & 60.21W & 56.62W \\
 \hline
 %Alibaba & 2048 & 60.21W & 56.62W \\
 \hline
  %Criteo Terabyte & 2048 & 62.8W & 58.47W \\
 %Criteo Terabyte & 4096 & 56.29W & 56.29W \\
 %Criteo Terabyte & 8192 & 62.86W & 56.73W \\
 %Criteo Terabyte & 16384 & 64.06W & 54.61W \\
\fi

%\input{body/discussion.tex}
\section{Related Work}

Training machine learning models is an important and heavily developed area of research. 
Optimizing training for deep neural networks training~\cite{largescalegputraining, dadiannao:micro:2014, cosmic:micro, gpipe, pipedream} has garnered most of the attention, whereas Recommender models have been under-researched.
%
%Prior work on recommender models with industry-scale datasets show that access patterns follow a Power or Zipfian distribution~\cite{yin2021ttrec} and thereby reaffirm our observation on hot embedding entries.

\niparagraph{Optimizations data layout through caching:}
Work in the past~\cite{informedcaching, ossdb} has delved into informed and domain-aware caching, which is highly pertinent to current applications, with their ever increasing requirement for compute and memory. 
In the deep learning realm, prior work~\cite{quiver} caches data on local SSD to eliminate slow reads from remote storage and employs hashing based techniques to incorporate thrashing-free strategies to efficiently utilize the shared cache. 
Instead, this work dives into the semantics of the training inputs observed by recommender models and offers compile time strategies to statistically ensure hot data is placed close to compute. 
\framework is able to fully exploit the coarse grained GPU based compute throughput without employing any dynamic hashing. 
Work in~\cite{entropy} and \cite{coordl} employ runtime techniques to improve memory, communication, and I/O resources for training and reduce data stall time, respectively. 
On the hardware side, works in ~\cite{bandana} propose techniques to store embedding tables in non-volatile memories and allocate a certain portion of DRAM for caching. 
This work, however, does not support GPU based training executions with replicated hot embeddings and does not deal with perceptive input pre-processing to reduce communication overheads. 
Recent work in ~\cite{mixedde, recnmp, compemd} has also proposed solutions to accelerate near-memory processing for embedding tables, but do not facilitate distributed training of entire recommender models using GPUs.

\if 0
\niparagraph{Efficient execution of tasks on GPUs:}
%
There has been a wide variety~\cite{concurrentqueries, hippogriff} of work across domains on optimizing GPU execution by improving its throughput and utilization.
%
Work in ~\cite{concurrentqueries} and ~\cite{mrshare} perform query executions by more efficiently by merging jobs into groups, thus reducing the setup overheads and increasing the throughput.
%
In this paper, we aim to reduce data transfer overheads and in turn increase GPU utilization by delving into recommender model properties.
%
Our work draws parallels to previous techniques by first understanding deep learning based recommender model's access patterns and devising a caching policy based on these observations.
\fi 

\niparagraph{Embedding parameter placement:} 
Works in~\cite{zhao2020distributed} offers a hierarchical parameter server that builds a distributed hash table across multiple GPUs. 
This work stores the working parameters close to computation, i.e, GPU, at runtime, albeit treats all embedding entries equally. 
Instead, \framework delves into the access pattern of each dataset and uses this information to store the highly accessed embedding entries in the GPU for the entirety of the training job.
Work in~\cite{acun2020understanding}, aims to understand the implications of different embedding table placements within an heterogeneous data-centre.  
However, none of the techniques leverage runtime access skew for their embedding table placement that can improve the overall training performance.

\niparagraph{Mitigating memory intensive training through compression, sparsity, and quantization:}
Past work has used compression~\cite{compressreco, gist, compressreco2},  sparsity~\cite{sparsemat}, and quantization~\cite{deepcompression} to reduce the overall memory footprint of machine learning models.
Prior work in \cite{nvopt}, optimizes training by modifying the model either through mixed-precision training or eliminating rare categorical variables to reduce the embedding table size. 
Even with these optimizations real dataset's entire embedding table cannot fit on a GPU.
Moreover, approaches that change the data representation and/or embedding tables, require accuracy re-validation across a variety of models and datasets.
\framework enables apropos utilization of memory hierarchy without employing overheads such as compression/decompression~\cite{hippogriff} and sparse operations.
\framework moreover performs full-precision training of the baseline model by leveraging the highly skewed access pattern for embedded tables and increase the throughput for hot embedding entries.
Nevertheless, our framework is orthogonal to the prior techniques and can be used in tandem with them to improve the memory efficiency even further.
%
%Nevertheless, we quantitatively compared NvOPT with FAE. Overall, \textbf{FAE is 1.48x faster than NvOPT} for the TeraByte Dataset. \framework reduces the per epoch training time from 105.98 mins to 71.58 mins (mini-batch size of 32K) on single V100 GPU as the most frequently access embedding tables are placed on the GPU memory.

\niparagraph{Distributed deep learning training:}
Data parallel training~\cite{dataparallel} forms the most common form of distributed training as it only requires synchronization after the gradients generated in backward pass of training. 
As models become bigger and bigger~\cite{tnlg, megatron}, model parallelism~\cite{modelparallel, gpipe} and pipeline parallelism~\cite{pipedream} are becoming common as they split a single model onto multiple devices.
Nonetheless, the techniques employed to automatically split the models~\cite{flexflow, deviceplacement:neurips}, offer model parallelism solutions to enable training of large model with size constrained by the accelerator memory capacity.
However, none of these techniques dive into the semantics of input data to perform an optimal split. This is because they are mainly suitable for DNNs. 
\section{Conclusions}
Recommendation models aim to learn user preferences and provide a targeted user experience by employing very large embedding tables.
Even though these tables often cannot fit on GPU memory, these models also comprise neural network layers that are well suited for GPUs. 
These contrasting requirements splits the training execution on CPUs (for memory capacity) and GPUs (for compute throughput).
Fortunately, for real-world data, we observe that embedding tables exhibit a skewed data access pattern. 
This can be attributed to certain training inputs (users and items) that are much more popular than the others.
This observation allows us to develop a comprehensive framework, namely \framework, that uses statistical techniques to quantify the hotness of embedding entries based on the input dataset. 
This hotness of embedding tables in turn allows the framework to optimally layout embedding so that the GPU memory is efficiently utilized to store highly accessed data close to the compute. %employ a GPU accelerated execution for inputs that only access these hot embedding entries.
To capture most of the performance benefits, \framework bundles hot inputs and cold inputs in separate mini-batches.
This helps \framework accelerate the hot mini-batch by executing the whole model on GPU and eliminate any CPU-GPU embedding data transfers.
The training for these hot inputs happens entirely on GPUs, thus reducing any CPU-GPU communication overhead between CPU-GPU and GPU-GPU from embedding and neural network layers.
Our experiments on DLRM and TBSM recommender models with real datasets show that \framework reduces the overall training time by \improvementavgxdlcpu and \improvementavgxdlgpu in comparison to XDL CPU-only and XDL CPU-GPU execution while maintaining baseline accuracy.

\balance
\section*{Acknowledgements}

This project is part of \textit{Phi Lab} at the University of British Columbia (UBC).
We express our thanks to the entire team of the Advanced Research Computing center at UBC~\cite{sockeye}.
We also thank the anonymous VLDB reviewers and Professor Arun Kumar for their invaluable feedback. 
This work was partially supported by the Natural Sciences and Engineering Research Council of Canada (NSERC) [funding reference number RGPIN-2019-05059]. 
This work is also research sponsored by Air Force Research Laboratory (AFRL) and Defense Advanced Research Project Agency (DARPA) under agreement number FA8650-20-2-7007.
The U.S. Government is authorized to reproduce and distribute reprints for Governmental purposes notwithstanding any copyright notation thereon. 

The views and conclusions contained herein are those of the authors and should not be interpreted as necessarily representing the official policies or endorsements, either expressed or implied, of Air Force Research Laboratory (AFRL), Defense Advanced Research Project Agency (DARPA), the U.S. Government, NSERC, the Canadian Government, Microsoft, or UBC.

\bibliographystyle{unsrtnat}
\bibliography{refs}

\end{document}